\documentclass[letterpaper,twocolumn,10pt]{article}
\usepackage{usenix2019_v3}

\usepackage{tikz}
\usepackage{amsmath}

\usepackage{colortbl}
\usepackage{listings}
\usepackage{tikz}
\usepackage{soul}
\usepackage{nicematrix}
\usepackage[normalem]{ulem}
\usepackage{tcolorbox}
\usepackage{diagbox}
\usepackage{multirow}
\usepackage{graphicx}
\usepackage{array}
\usepackage{paralist}
\usepackage{enumitem}
\usepackage{wrapfig}
\usepackage{calc}
\usepackage{makecell}

\usetikzlibrary{arrows}
\usetikzlibrary{backgrounds}
\usetikzlibrary{fit}
\usetikzlibrary{calc}
\usetikzlibrary{positioning}
\usetikzlibrary{shapes}

\pgfdeclarelayer{background}
\pgfdeclarelayer{foreground}
\pgfsetlayers{background,main,foreground}

\lstdefinelanguage[RISC-V]{Assembler}
{
  alsoletter={.}, 
  alsodigit={0x}, 
  morekeywords=[1]{ 
    lb, lh, lw, lbu, lhu,
    sb, sh, sw,
    sll, slli, srl, srli, sra, srai,
    add, addi, sub, lui, auipc,
    xor, xori, or, ori, and, andi,
    slt, slti, sltu, sltiu,
    beq, bne, blt, bge, bltu, bgeu,
    j, jr, jal, jalr, ret,
    scall, break, nop, csrwi, mv, ld, sd
  },
  morekeywords=[2]{ 
    .align, .ascii, .asciiz, .byte, .data, .double, .extern,
    .float, .globl, .half, .kdata, .ktext, .set, .space, .text, .word
  },
  morekeywords=[3]{ 
    zero, ra, sp, gp, tp, s0, fp,
    t0, t1, t2, t3, t4, t5, t6,
    s1, s2, s3, s4, s5, s6, s7, s8, s9, s10, s11,
    a0, a1, a2, a3, a4, a5, a6, a7,
    ft0, ft1, ft2, ft3, ft4, ft5, ft6, ft7,
    fs0, fs1, fs2, fs3, fs4, fs5, fs6, fs7, fs8, fs9, fs10, fs11,
    fa0, fa1, fa2, fa3, fa4, fa5, fa6, fa7
  },
  morecomment=[l]{;},   
  morecomment=[l]{//},   
  morecomment=[l]{\#},  
  morestring=[b]",      
  morestring=[b]'       
}

\makeatletter
\newcommand{\code}[1]{\protect\lstinline[language={[RISC-V]Assembler},basicstyle=\ttfamily]|#1|}
\makeatother

\definecolor{mauve}{rgb}{0.58,0,0.82}
\lstdefinestyle{myCstyle}{
  language=C,
  frame=none,
  basicstyle=\tiny\ttfamily,
  keywordstyle=\color{blue},
  keywordstyle=[2]\color{red},
  stringstyle=\color{red},
  commentstyle=\color{green},
  morekeywords=[2]{BURST_ON, BURST_OFF},
  showstringspaces=false
}

\definecolor{LightRed}{rgb}{1,0.6,0.6}
\definecolor{LightRedbis}{rgb}{0.9,0.9,0.9}
\definecolor{mygreen}{rgb}{0.1, 0.5, 0.1}
\definecolor{myblue}{RGB}{220, 231, 250}
\newcommand*\circled[1]{\tikz[baseline=(char.base)]{
            \node[shape=circle,draw,inner sep=1pt] (char) {#1};}}

\definecolor{lightred}{RGB}{255,200,200}
\definecolor{lightorange}{RGB}{255,255,200}
\definecolor{lightgreen}{RGB}{200,255,200}
\newcommand{\red}[1]{\cellcolor{lightred}#1}
\newcommand{\orange}[1]{\cellcolor{lightorange}#1}
\newcommand{\green}[1]{\cellcolor{lightgreen}#1}

\usepackage{array} 

\usepackage{stmaryrd}  
\usepackage{amssymb}
\definecolor{semanticsred}{RGB}{255,0,0}    
\definecolor{semanticsredf}{RGB}{255,150,150}    
\definecolor{semanticsgreen}{RGB}{0,128,128} 

\newcommand{\uarch}[3][]{{\color{semanticsred}\{\!\!|}#2{\color{semanticsred}|\!\!\}_{#1}^{#3}}}
\newcommand{\uarchf}[3][]{{\color{semanticsredf}\{\!\!|}#2{\color{semanticsredf}|\!\!\}_{#1}^{#3}}}

\newcommand{\contract}[3][]{{\color{semanticsgreen}\llbracket}#2{\color{semanticsgreen}\rrbracket_{#1}^{#3}}}

\newcommand{\hw}[1]{{\color{semanticsred}#1}}
\newcommand{\sw}[1]{{\color{semanticsgreen}#1}}

\newif\ifdraft 
\draftfalse

\newcommand{\MemExePipeline}{MemExePipeline}
\newcommand{\MemRS}{MemRS}

\newcommand{\prob}{$\mathcal{P}$}

\newcommand{\ydistance}{1}

\newcommand{\xdistance}{1}

\begin{document}

\date{}

\title{Citadel: Simple Spectre-Safe Isolation For Real-World Programs \\ That Share Memory}

\author{
 {\rm Jules Drean}\\
 MIT CSAIL
\and
{\rm Miguel Gomez-Garcia}\\
MIT CSAIL
\and
{\rm Fisher Jepsen}\\
MIT CSAIL
\and
{\rm Thomas Bourgeat}\\
EPFL
\and
{\rm Srinivas Devadas}\\
MIT CSAIL
 } 

\maketitle

\begin{abstract}
Transient execution side-channel attacks, such as Spectre, have been shown to break almost all isolation primitives.
We introduce a new security property we call \emph{relaxed microarchitectural isolation} (RMI) that allows \emph{sensitive} programs that are \emph{not-constant-time} to \emph{share memory} with an attacker while restricting the information leakage to that of non-speculative execution.
Although this type of speculative security property is typically challenging to enforce, we show that we can leverage the \emph{enclave} setup to achieve it.
In particular, we use microarchitectural isolation to restrict attacker's observations in conjunction with straightforward hardware mechanisms to limit speculation.
This new design point presents a compelling trade-off between security, usability, and performance, making it possible to efficiently enforce RMI for \emph{any} program.

We demonstrate our approach by implementing and evaluating two simple defense mechanisms that satisfy RMI: (1) \emph{Safe mode}, which disables speculative accesses to shared memory, and (2) \emph{Burst mode}, a localized performance optimization that requires simple program analysis on small code snippets.
Our end-to-end prototype, Citadel, consists of an FPGA-based multicore processor that boots Linux and runs secure applications, including cryptographic libraries and private inference, with less than 5\% performance overhead.
\end{abstract}

\section{Introduction}
\begin{figure}[t]
    \centering
    \includegraphics[width=\columnwidth]{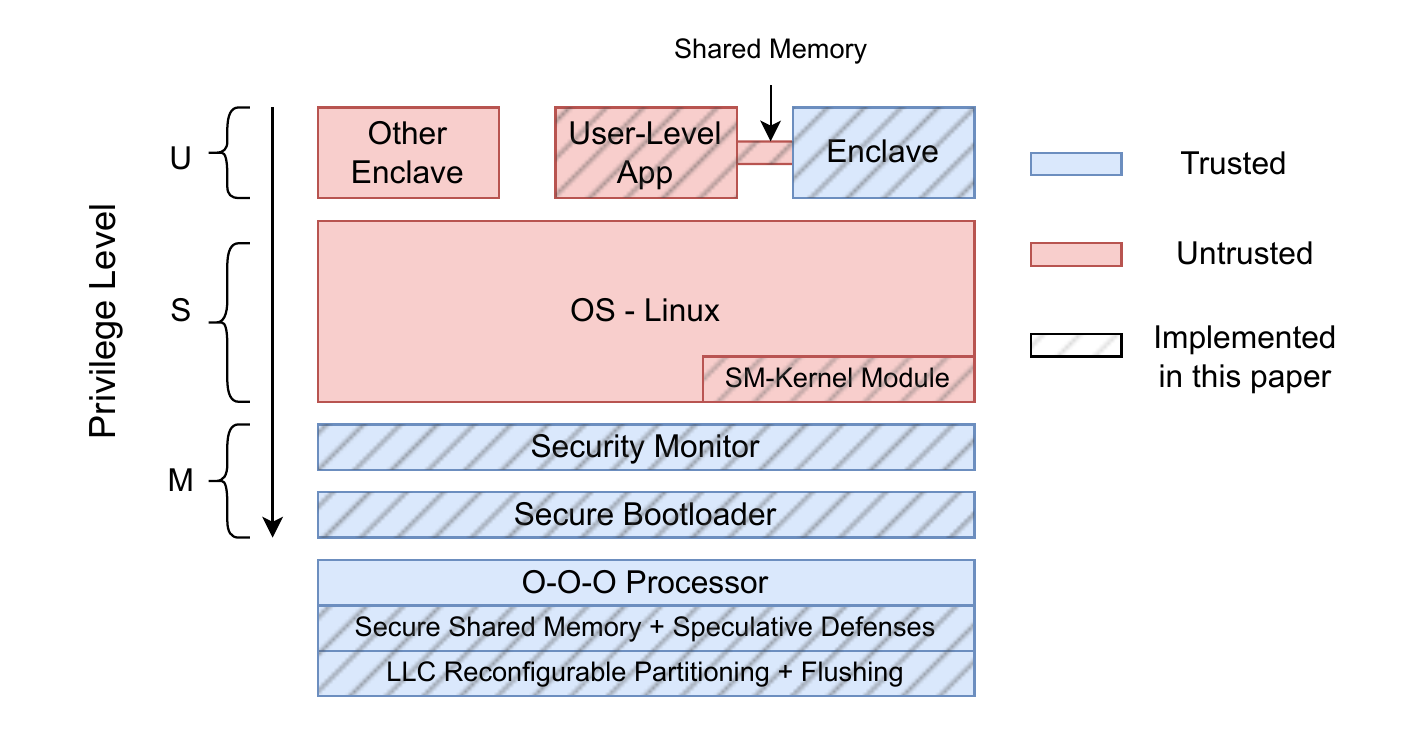}
    \caption{Citadel's Computing Stack.}
    \label{fig:computing_stack}
    \end{figure}

Citadel proposes a simple and efficient solution to an important problem: ``\emph{How to defend against transient execution side channels for a program that 1) touches secrets, 2) is not written in constant-time, and 3) shares some memory with an attacker}''. 
We call this problem \prob{}. 
It represents a common case for \emph{security-sensitive} programs in the real world which do \emph{not} aim for leakage-free execution times.
For example, consider a private inference service~\cite{ApplePrivateCloud} that needs to isolate the machine learning model and private user requests from the rest of the system, while still accessing the network, a GPU, and communicating with other processes in charge of access control and business logic.

Transient execution side channel (TES) attacks such as Spectre \cite{kocher2018spectre} combine speculative execution and microarchitectural side channels to exfiltrate secrets.
Among the numerous Spectre variants, those exploiting cache side channels are some of the most powerful.
These attacks only require an attacker and the victim to share \emph{any} address space (e.g., through shared memory) to make it possible to construct ``universal read gadgets,'' exposing all the victim's private memory.
They are also particularly challenging to mitigate and circumvent almost all hardware and software isolation mechanisms.
For example, in a scenario like serverless computing, these attacks can compromise the isolation between containers and virtual machines, or enable privilege escalation~\cite{weissman2023microarchitectural}.

Trusted execution environments (TEEs) or secure enclaves \cite{costan2016sanctum, ferraiuolo2017komodo, lee2020keystone,  bahmani2021cure, feng2021scalable, li2022design, cheng2024intel, costan2016intel, mckeen2013sgx, brasser2019sanctuary, sev2020strengthening, trustzone04, pinto2019demystifying} are some of the most ambitious isolation primitives.
These hardware mechanisms aim to completely isolate sensitive processes even in the presence of a privileged software attacker, such as a malicious operating system (OS) or hypervisor (for example, see Figure~\ref{fig:computing_stack}).
However, most of these platforms have been proven vulnerable to TES attacks~\cite{koruyeh2018spectre, chen2019sgxpectre, ryan2019hardware, van2020lvi, van2018foreshadow, van2020sgaxe, gotzfried2017cache, brasser2017software, gras2018translation, evtyushkin2018branchscope, aciiccmez2006predicting, wang2014timing, koruyeh2018spectre}.
On commercial enclave platforms such as Intel SGX~\cite{costan2016intel} or AMD SEV~\cite{sev2020strengthening}, defending against TES is still an open problem, and all responsibility is pushed to program developers.
Countermeasures rely on the placement of fences and special instructions in the binary (see Section \ref{sec:softwaremitigations}), a challenging process that needs to balance preservation of performance and complexity of code analysis, often achieving unclear security guarantees.

One academic work, MI6~\cite{bourgeat2019mi6}, looks at securing enclave programs in the presence of TES through complete microarchitectural isolation.
However, they took the approach to the extreme by disabling shared memory altogether.
This significantly reduces the usability and applicability of enclaves, in contrast to commercial enclave platforms and other isolation primitives such as containers or trusted VMs, which all rely heavily on shared memory to implement many important functionalities, such as input passing, I/O, device virtualization, or micro-services chaining (see Section~\ref{sec:motivsharedmem}).

Beyond enclave-specific solutions, other works have explored software and hardware defense mechanisms against TES attacks (see Section \ref{sec:relatedwork}).
Each of these mechanisms addresses a slightly different trade-off between security, performance, and use cases (see Sections~\ref{sec:beyondsandboxing}).
Within that design space, Citadel and its simple mechanisms stand out as well equipped to efficiently address \prob{}.
\\

\textbf{This paper.}
We start by defining a new speculative security property---\emph{relaxed microarchitectural isolation} (RMI)---that allows \emph{non-constant-time} programs that \emph{touch secrets} to \emph{share memory} (i.e. to address \prob) while providing precise guarantees on information leaked to an attacker.
For a given defense scheme, RMI is achieved if information leakage through the microarchitecture is equivalent to the trace of \emph{shared memory accesses} when executing the program non-speculatively (in-order, one instruction at a time).
We use the formalism from \emph{hardware-software contracts for security}~\cite{guarnieri2020spectector,guarnieri2021hardware, mosier2022axiomatic, cauligi2022sok} to compare and relate RMI with other existing speculative security properties (see Figure~\ref{figure:guarantees}.
We show that RMI is equivalent to a \emph{weak/simple} contract, which makes it possible to secure a very large class of programs despite enforcing a very strong security property under a strong threat model.

We describe RMI in the context of shared memory between an enclave and the OS.
However, our work also applies to \emph{private} shared memory between programs and can help compose code following other non-interference properties such as speculative constant time~\cite{guarnieri2020spectector}.
Our work also extends to other types of isolation mechanisms and TEEs, such as containers or trusted VMs.

To enforce RMI, we design Citadel using two strategies: 1) we use microarchitectural isolation to weaken attacker capabilities and satisfy a weaker leakage model.
2) we use simple hardware mechanisms to constrain speculation in order to enforce simple execution models.
Because we focus on programs that are \emph{not} implemented using constant time and do \emph{not} aim for leakage-free execution times, our threat model allows for some leakage of coarse grain timing information (see Section \ref{sec:execution_time_SC}).
This offers a compelling trade-off between security and enabling \emph{simpler} and \emph{more efficient} mechanisms that can secure a \emph{large class} of enclave programs against Spectre-like attacks.

Starting from a design that enforces strong microarchitectural isolation, we show how to enable shared memory while limiting the attacker observations to the trace of (speculative) accesses to shared memory.
This requires being careful with the core-private state that can now be shared between private and shared memory.
For example, we highlight how an attacker can observe the private state of L1 by exploiting a new side channel that leverages the cache coherence protocol.

Once attacker observations have been restricted, we present two \emph{simple} approaches to limit speculative leakage: (1) {\bf Safe mode} is our default execution mode and prevents speculative shared memory accesses entirely, making it possible to enforce RMI for \emph{any} program, (2) {\bf Burst mode} can be enabled \emph{locally} to allow some speculation on shared memory accesses and improve overall performance. 

The Burst mode microarchitecture enforces a slightly stronger contract compared to Safe mode (more speculative control flows are considered public).
As a result, Burst needs to be supplemented by a static analysis tool that identifies programs that satisfy the correct \emph{relative non-interference} property.
Nevertheless, the property is simple enough that programmers can easily reason about it and can be checked automatically using some simple tool.
We use a contract formalism to show that the resulting hardware-software defense satisfies RMI.
We show how code analysis can be simplified by restricting it to small code snippets, circumventing the usual limitations of automated tools.
Note that code annotation and analysis are only required to improve \emph{performance} and that all programs are secure as is without manual annotation.
Code can be written to use the two modes simultaneously and can be safely composed or reused without the need for additional analysis. 
\\

\textbf{Our prototype.} Citadel, is an out-of-order RISC-V 2-core processor synthesized for an AWS F1 FPGA.
We apply our design principles and implement all mechanisms at the microarchitectural level.
We restrict the attacker's observations through L1-cache bypass and TLB / translation cache tagging.
We also implement a configurable cache partitioning mechanism, improving performance for both enclaves and unprotected processes.
We implement \emph{Safe} mode and \emph{Burst} modes with minimal hardware overhead (<+1\% LUTs and FFs).
To evaluate our prototype, we implement a comprehensive software infrastructure that includes a bootloader, a security monitor, a kernel module, and an end-to-end attestation mechanism.
Citadel boots (untrusted) Linux and enables user-level applications to interact safely with secure enclaves.
For unprotected software, we measure negligible overhead ($\simeq0.1\%$).
To demonstrate the platform's versatility, we port three security-sensitive programs: a lightweight Python runtime, a cryptographic library, and a private ML inference program.
We show that very little implementation effort is required to port the applications to Citadel: Less than $200$ new LOC and minimal automated code analysis for performance overheads under $5\%$.
Figure~\ref{fig:computing_stack} provides an overview of our implementation work.
We open source our hardware and software infrastructure at \url{https://github.com/mit-enclaves/citadel}.

\smallskip\noindent
In summary, our contributions include the following:
\begin{itemize}[leftmargin=*, align=left, itemsep=0ex, topsep=0ex]
\item 
    We present the first microarchitecturally-isolated TEE with Spectre-safe shared memory;
\item 
    We introduce the notion of \emph{relaxed microarchitectural isolation}, a new security property that allows non-constant-time, sensitive programs to share memory while thwarting TES attacks;
\item 
    We demonstrate that microarchitectural isolation and simple hardware mechanisms for controlled speculation make it possible to effectively defend against most TES attacks;
\item 
    We implement and evaluate two simple mechanisms---Safe mode and Burst mode---that achieve RMI for any program while offering the option to balance performance and the need for simple program analysis;
\item 
    We show how to compose code with different speculative security properties and that program analysis can be constrained to small code snippets, circumventing the usual scaling issues of automated tools;
\item
    We use formalism from hardware-software contracts for security to compare RMI with other security property and show that Citadel enforces it;
\item
    We show that our approach allows for minimal hardware overhead (<1\% for Safe and Burst mode) and minimal performance overhead (<5\% on our benchmarks);
\end{itemize}

\bigskip
\noindent\textbf{Disclaimer.} Citadel is a \emph{simple} and \emph{efficient} solution to an important problem.
We do not claim the invention of the simple hardware mechanisms we compose to achieve security.
On the contrary, we aim to highlight that complex mechanisms are not necessary here.
Similarly, most of the insights we present might be considered folklore by some readers.
We show how all these elements are tightly integrated to create a novel point in the design space at the intersection of TEEs and TES defense mechanisms, presenting a compelling trade-off between performance, usability, and security.
\section{Background}

\noindent\textbf{Security Monitor:}
\label{sec:securitymonitor}
In many enclave platforms, the security monitor (SM) \cite{costan2016sanctum, lebedev2019sanctorum} is a small (9K LOC in our prototype), trusted piece of software running at a higher privilege mode than the hypervisor or the OS (see Figure \ref{fig:computing_stack}).
Its role is to link low-level invariants exposed by the hardware (e.g., if a specific memory access is authorized), to the high-level security policies defined by the platform (e.g., which enclave can access which memory regions) and to invoke the right microarchitectural cleanup routine.
\\ 

\noindent\textbf{Microarchitectural Timing Side Channels:}
Microarchitectural side channels enable information leakage between security domains by exploiting shared microarchitectural structures. These include memory caches \cite{gotzfried2017cache, brasser2017software}, translation look-aside buffers (TLBs) \cite{gras2018translation}, branch predictors \cite{evtyushkin2018branchscope, aciiccmez2006predicting}, DRAM controllers \cite{wang2014timing}, or any other shared microarchitecture. Attackers typically rely on event timing to indirectly observe microarchitectural state, as it cannot be directly accessed through the instruction set architecture (ISA).
For example, cached data access is faster than main memory access. By manipulating shared microarchitectural state and observing changes due to victim activity, attackers can infer sensitive information about the victim's execution, such as accessed addresses or branch directions.
\\ 

\noindent\textbf{Transient Execution Side Channels (TES):}
\label{sec:specandside}
Speculative or transient execution, a key performance feature in modern processors, complicates security boundary reasoning.
Attackers can manipulate microarchitectural structures that orchestrate speculation (like branch predictors) to trigger speculative execution of code snippets that bypass normal security checks.
Although these erroneous executions are eventually squashed, they alter microarchitectural state (e.g., cache states), potentially leaking sensitive information through side channels.
This class of attacks, known as Spectre, was introduced in Kocher et al. \cite{kocher2018spectre} and has since spawned numerous variants.
\\

\noindent\textbf{Hardware Software Contract for Security:}
Recent work \cite{guarnieri2021hardware, mosier2022axiomatic,cauligi2022sok} has focused on defining formal foundations for speculative security properties using \emph{hardware-software contracts}.
Contract semantics consist of two components: a \emph{leakage model} (\sw{leak}) and an \emph{execution model} (\sw{exec}) and are represented using the notation $\contract[leak]{\cdot}{exec}$.
The leakage model describes what \emph{architectural} state is made public by the software (e.g. execution trace and trace of memory accesses for the \sw{ct} model).
The execution model, on the other hand, models the ``safe'' speculative behavior i.e. which speculative control flows are considered \emph{public} by the software.
A defense mechanism, represented by the \emph{attacker model} $\uarch[]{\cdot}{}$ (formally a \emph{projection} of the microarchitectural state), satisfies a \emph{contract} $\contract[]{\cdot}{}$ if it does not leak more information than what the contract specifies as public.
We write $\uarch[]{\cdot}{} \vdash \contract[]{\cdot}{}$.
Although the software does not directly satisfy contracts per say, it can verify \emph{non-interference} properties with respect to contracts.
For a program $p$, \emph{direct non-interference} with respect to contract $\contract[]{\cdot}{}$ means the program will not make any \emph{secret} information public under that contract.
We write this as $p \vdash NI(\contract[]{\cdot}{})$ and say $p$ satisfies $\contract[]{\cdot}{}$ for simplicity.
Additionally, \emph{relative non-interference} from $\contract[]{\cdot}{A}$ to $\contract[]{\cdot}{B}$, means that it will not expose more information under $\contract[]{\cdot}{B}$ than what is made public by $\contract[]{\cdot}{A}$.
We write $p \vdash NI(\contract[]{\cdot}{A} \Rightarrow \contract[]{\cdot}{B})$.
Detailed definitions are recalled in appendix~\ref{app:hwsfcontract}.

\section{Motivation}

\subsection{Utility of Shared Memory}
\label{sec:motivsharedmem}

In all commercial enclave platforms, shared memory serves as the primary communication primitive for crucial mechanisms including input/output/message passing, networking \cite{costan2016intel, kaplan2016amd}, or access to large public data structures that exceed private memory capacity (e.g., public neural network for private inference as in Section \ref{sec:evalprivateinference}).
Other isolation primitives like containers and trusted VMs also dependent on shared memory to efficiently chain micro-services\cite{sabbioni2021shared, qi2022spright} or to implement mechanisms like VirtIO \cite{IntelVirtIO}.
Alternative approaches such as message passing or DMA copy mechanisms are limited to passing small inputs or to sequential memory accesses, making it overly complex to do tasks such as accessing a linked list of inputs or doing random memory accesses, for example, accessing a database using ORAM \cite{signalcontact}.
For instance, MI6 briefly mentions a mailbox mechanism which demands application rewrites, limits message sizes to a few bytes, and incurs substantial overheads due to two data copies and two full microarchitectural context switches for each message.

\begin{figure}[t]
    \centering
    \includegraphics[width=\columnwidth]{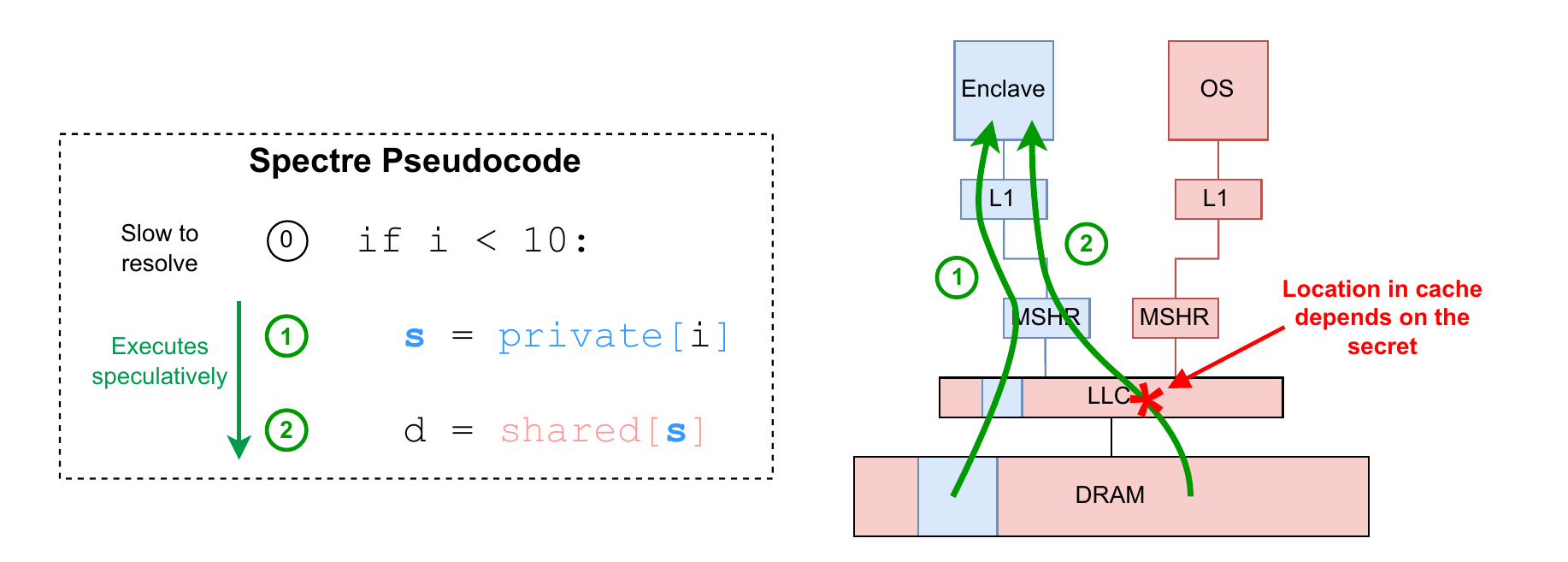}
    \caption{Universal Read Gadget Through Shared Memory}
    \label{fig:specter_attack}
\end{figure}

\subsection{Universal Read Gadget Through Shared Memory}
\label{sec:attackandexisting}
Shared memory enables LLC side channels and Spectre-like attacks. 
We demonstrate a proof-of-concept variant Spectre-PHT or v1 \cite{kocher2018spectre} attack on our insecure FPGA baseline, exfiltrating \emph{arbitrary} private enclave memory via the shared LLC side channel (see Figure \ref{fig:specter_attack}).
Note that this insecure baseline is also vulnerable to other Spectre variants like BTB, RSB or STL \cite{canella2019systematic, wiebing2024inspectre, barberis2022branch, koruyeh2018spectre, maisuradze2018ret2spec, horn2018speculative}.
The victim enclave exposes an API allowing access to a public array, with out-of-bounds accesses protected by a branch.
The value fetched from the first array is used to access a second array located in shared memory, effectively leaking the first value through the cache side channel.
This creates a ``universal read gadget,'' commonly found in code \cite{wiebing2024inspectre, johannesmeyer2022kasper, qi2021spectaint, canella2019systematic}.
By training the branch predictor, a malicious OS can perform speculative out-of-bounds accesses, leaking arbitrary enclave secrets.
In particular, shared-memory content (encrypted or not) is irrelevant, as information is leaked through which address is accessed.
The code for our attack is available in our repository.
We were unable to successfully mount the attack on our final secure design.

\subsection{Going Beyond the Sandboxing and Constant-Time Scenario}
\label{sec:beyondsandboxing}
\noindent\textbf{Sandboxing Scenario}
Many existing defense mechanisms against TES attacks such as STT~\cite{yu2019speculative, jauch2023secure} or SpecShield~\cite{barber2019specshield} try to enforce ``speculative memory safety'' or ``speculative isolation'', that is, to enforce the \emph{sandbox isolation model}~\cite{cheang2019formal, narayan2021swivel, qi2021spectaint, shen2018restricting, jenkins2020ghostbusting, oleksenko2020specfuzz, kirzner2021analysis, yu2020speculative, narayan2023going, behrens2020efficiently}.
Sandboxing is an isolation primitive that aims to protect the rest of a system from a malicious program.
This is a \emph{much weaker} threat model than enclaves that aim to \emph{also protect the isolated program}.
Specifically, the mechanisms in this scenario often assume that the architectural state of the isolated program is \emph{public} and is \emph{not} protected against TES attacks.
As a result, these mechanisms are not the best candidate building blocks for isolating \emph{sensitive} programs that \emph{touch secrets}, therefore to address \prob{}.
\\

\noindent\textbf{Constant Time Scenario}
Other proposals, such as Prospect \cite{daniel2023prospect} or SPT \cite{choudhary2021speculative} and others~\cite{barthe2021high, cauligi2020constant, daniel2021hunting, guanciale2020inspectre, guarnieri2020spectector, patrignani2021exorcising, vassena2021automatically} only target \emph{constant-time programs}.
This is a very important problem to solve but only concerns a subset of programs, namely cryptographic algorithms written in a very specific constant time fashion~\cite{almeida2016verifying}.
This also makes these mechanisms overkill to specifically address \prob{}.
No programmer will rewrite operating systems (in the case of trusted VM), Python libraries, data analysis pipelines, or business logic using constant-time programming.
Citadel focuses on defining a speculative security property and building defense mechanisms for the majority of \emph{security-sensitive} programs, which are \emph{not} implemented using constant time and do \emph{not} aim for leakage-free execution times (see Section \ref{sec:execution_time_SC}).
\\

\noindent\textbf{Speculative Non-Interference is Too Restrictive}
\label{sec:limitationsspecsec}
Existing hardware defenses studied under the hardware-contract framework only target the two above scenarios.
They use one of two \emph{strong} leakage models, each defining a lot of architectural state as public.
The \emph{constant-time} model (\sw{ct} for short) makes public the trace of the \texttt{pc} (i.e. the control flow) and the trace of all memory accesses (but not their values).
The \emph{architectural} or \emph{sandboxing} model (\sw{arch}) extends the \sw{ct} model to include the \emph{values} of elements loaded from memory. 
Figure~\ref{figure:guarantees} shows how the resulting contracts relate to each other.
Working with these strong leakage models strongly restricts the class of programs that can be run without exposing secrets (i.e., programs that satisfy those contracts), and only works in the above-mentioned scenarios.
This makes these leakage models, along with any associated speculative security properties, too strong to address \prob{} efficiently.
This includes properties such as speculative constant time, (weak) speculative non-interference, and others.
A summary of all these properties can be found in the SoK by Cauligi et. al.~\cite{cauligi2022sok}.
\section{Threat Model and Security Property}
\label{sec:threatmodel}

\subsection{Attacker Capabilities}
We follow the usual attacker model for TEEs.
We assume a privileged software attacker co-located on the same machine as the victim enclave.
The attacker has compromised the majority of the software stack, including other enclaves and the OS, except for a limited trusted code base (TCB) including the secure bootloader, the SM (together 15K lines of code (LOC)) and the victim program running inside the enclave (see Figure \ref{fig:computing_stack}).
Note that the TCB may contain Spectre gadgets.
The attacker also has control over the contents of shared memory and may control some of the enclave's inputs.
For the rest of this paper, we refer to any code outside the TCB as the OS.
Attackers can exploit the timing of shared microarchitectural events to mount TES attacks.

\subsection{Attacks Considered Out-of-Scope}
\noindent\textbf{Execution-Time Side Channel}
\label{sec:execution_time_SC}
The time an enclave takes to execute might depend on a secret which can be exploited by an attacker.
This includes the timing between any observable (micro)architectural events like syscalls or accesses to shared memory.
We refer to these attacks as exploiting "execution time".
These attacks can only be addressed for constant-time programs \cite{daniel2023prospect, daniel2021hunting, schwarz2020context, choudhary2021speculative, yu2019speculative}. 
To defend against execution-time attacks for this restricted class of programs, Citadel offers the option to disable speculation as we do in the SM (we evaluate a +882\% overhead on a crypto library -- see Section \ref{sec:evalcrypto}).
However, because we mostly want to protect programs that are \textit{not} constant time,
we follow previous work \cite{bourgeat2019mi6, khasawneh2019safespec, kiriansky2018dawg, sakalis2019efficient, li2019conditional, ainsworth2020muontrap, saileshwar2019cleanupspec, wang2019oo7} and consider execution-time attacks out of scope.
These attacks are generally considered lower risk due to lower bandwidth, higher susceptibility to noise, and increased difficulty compared to other side channels. 
For example, NetSpectre \cite{schwarz2019netspectre} can only extract 15 bits per hour using methods applicable to Citadel compared to rates more than 1MB/s for LLC-based side channels \cite{liu2015last}.

\noindent\textbf{Others}
We do not consider physical attacks (e.g., noise, electromagnetic field) or side channels that exploit the physical layer (e.g. power side channels). 
We do not consider Rowhammer attacks.
We are concerned about the privacy of the enclave and do not consider denial-of-service.
As expressed by our contract, we also do not protect enclave programs that explicitly leak their own secrets, including through shared memory.

\subsection{New Building Blocks for Contracts}

\noindent\textbf{Leakage Models}
We define \sw{shm}, a new \emph{weak} leakage model that only represents the trace of accesses to \emph{shared} memory.

\noindent\textbf{Execution Models}
The weakest execution model is \sw{seq}, which represents non-speculative, or sequential execution (one instruction at a time and in-order).
We also have the strongest execution model \sw{spec} which models speculative execution with the following predictors: the pattern history table (PHT), the branch target buffer (BTB) and the return stack buffer (RSB).
We also define \sw{stl} that models straight-line speculation~\cite{straightlinespecARM} where branches are always speculated as non-taken (static branch predictor).
\\

\subsection{Relaxed Microarchitectural Isolation (RMI)}

Given our threat model, we propose a new security property that allow programs to safely share memory.

 \noindent\fbox{%
    \parbox{\dimexpr\columnwidth-2\fboxsep-2\fboxrule\relax}{%
        For a given program, \emph{relaxed microarchitectural isolation} is achieved if information leakage through the microarchitecture is equivalent to the trace of shared memory accesses when running the program non-speculatively (one-instruction-at-a-time and in-order).
    }%
}
That means, for a defense mechanism $\uarch[]{\cdot}{}$, enforcing RMI is exactly equivalent to satisfying $\contract[shm]{\cdot}{seq}$.
Figure \ref{figure:guarantees}, represents how RMI relates to contracts referenced in previous work.
\\

\noindent\textbf{Our goal} is to design microarchitecture that enforces RMI for a large class of programs.
Our design strategy is two-fold.

\begin{enumerate}[leftmargin=*, align=left, itemsep=0ex, topsep=0ex]
    \item Restricting Attacker Observations (HW): we use microarchitectural isolation to restrict the attacker observations to the trace of (speculative) accesses to shared memory.
    \item Restricting Speculative Leakage (HW+SW): we use a combination of mechanisms for controlled speculation and program analysis to ensure these observation traces do not leak more information than if the program was run non-speculatively.
\end{enumerate}

\renewcommand{\ydistance}{0.8}
\renewcommand{\xdistance}{0.75}

\begin{figure}[t!]
    \centering

    \footnotesize
    
    \begin{tikzpicture}[->,>=stealth',shorten >=1pt,auto, semithick]


    \node[fill=none,draw=none, shape = rectangle, rounded corners, inner sep=0pt, outer sep=0pt, minimum height = 1pt, minimum width = 1pt]  (seqCt) at (0,0) {$\contract[ct]{\cdot}{seq}$};

    \node[fill=none,draw=none, shape = rectangle, rounded corners, inner sep=0pt, outer sep=0pt, minimum height = 1pt, minimum width = 1pt]  (specArch) at ($(seqCt) + (-2*\xdistance,-2.5*\ydistance)$) {$\contract[arch]{\cdot}{spec}$};

    \node[fill=none,draw=none, shape = rectangle, rounded corners, inner sep=0pt, outer sep=0pt, minimum height = 16pt, minimum width = 1pt]  (seqArch) at ($(seqCt) + (-2*\xdistance,-0.5*\ydistance)$) {$\contract[arch]{\cdot}{seq}$};

    \node[fill=none,draw=none, shape = rectangle, rounded corners, inner sep=0pt, outer sep=0pt, minimum height = 16pt, minimum width = 1pt]  (specCt) at ($(seqCt) + (0*\xdistance,-2*\ydistance)$) {$\contract[ct]{\cdot}{spec}$};



     \node[fill=none,draw=none, shape = rectangle, rounded corners, inner sep=0pt, outer sep=0pt, minimum height = 16pt, minimum width = 1pt]  (seqMem) at ($(seqCt) + (2*\xdistance,0.5*\ydistance)$) {$\contract[mem]{\cdot}{seq}$};

     \node[fill=none,draw=none, shape = rectangle, rounded corners, inner sep=0pt, outer sep=0pt, minimum height = 16pt, minimum width = 1pt]  (seqShm) at ($(seqMem) + (2*\xdistance,0.5*\ydistance)$) {$\contract[shm]{\cdot}{seq}$};


     \node[fill=none,draw=none, shape = rectangle, rounded corners, inner sep=0pt, outer sep=0pt, minimum height = 16pt, minimum width = 1pt]  (specShm) at ($(specCt) + (4*\xdistance,1*\ydistance)$) {$\contract[shm]{\cdot}{spec}$};

      \node[fill=none,draw=none, shape = rectangle, rounded corners, inner sep=0pt, outer sep=0pt, minimum height = 16pt, minimum width = 1pt]  (stlShm) at ($(specShm) + (0*\xdistance,1*\ydistance)$) {$\contract[shm]{\cdot}{stl}$};


      \node[fill=none,draw=none, shape = rectangle, rounded corners, inner sep=0pt, outer sep=0pt, minimum height = 16pt, minimum width = 1pt]  (bot) at ($(seqShm) + (2*\xdistance,0*\ydistance)$) {$\contract[\bot]{\cdot}{}$};


    \path (seqArch) edge[] node[left] {} (seqCt);
    \path (specArch) edge[] node[left] {} (seqArch);
    \path (specArch) edge[] node[left] {} (specCt);
    \path (specCt) edge[] node[left] {} (seqCt);
    \path (seqCt) edge[] node[left] {} (seqMem);
    \path (specCt) edge[] node[left] {} (specShm);
    \path (seqMem) edge[] node[left] {} (seqShm);
    \path (seqShm) edge[] node[left] {} (bot);
    \path (specShm) edge[] node[left] {} (stlShm);
    \path (stlShm) edge[] node[left] {} (seqShm);

    
    \begin{scope}[on background layer]

        \node[draw=gray!50,dashed, fill=none, fill opacity=0.5, rounded corners, rotate fit=0, minimum height=0.1em, fit=(seqCt)](FitCt) {};
        \node  (label1) at ($(FitCt.north) + (0,0.25)$) {$\uarchf[pro]{\cdot}{}$};
        
        \node[draw=gray!50,dashed, fill=none, fill opacity=0.5, rounded corners, rotate fit=-36, minimum height=0.1em, fit=(seqArch) (specCt)](FitTt) {};
        \node  (label2) at ($(FitTt.east) + (0.5,0)$) {$\uarchf[tt]{\cdot}{}$};
        


        \node[draw=red!100,dashed, fill=none, fill opacity=0.5, rounded corners, rotate fit=0, minimum height=0.1em, fit=(seqShm)](FitSafe) {};
        \node  (label5) at ($(FitSafe.north) + (0,0.25)$) {$\uarch[safe]{\cdot}{} \hw{/} \uarch[burst]{\cdot}{sta}$};

        \node[draw=gray!50,dashed, fill=none, fill opacity=0.5, rounded corners, rotate fit=0, minimum height=0.1em, fit=(stlShm)](FitBurstRaw) {};
        \node  (label7) at ($(FitBurstRaw.east) + (0.5,0)$) {$\uarchf[burst]{\cdot}{\top}$};


        \node[draw=gray!50,dashed, fill=none, fill opacity=0.5, rounded corners, rotate fit=0, minimum height=0.1em, fit=(bot)](FitMI6) {};
        \node  (label9) at ($(FitMI6.north) + (0,0.25)$) {$\uarchf[mi6]{\cdot}{}$};
    \end{scope}

    \end{tikzpicture}
    \caption{Security guarantees of secure-speculation mechanisms. 
    Hardware semantics of Citadel' defense modes, Safe ($\uarch[safe]{\cdot}{}$) and Burst ($\uarch[burst]{\cdot}{sta}$), both satisfy RMI ($\contract[shm]{\cdot}{seq}$).
    We also represented other relevant hardware semantics such as $\uarch[burst]{\cdot}{\top}$ which represents the hardware mechanisms of Burst without the associated static code analysis (see Section \ref{sec:burstsecanalysis}), and some previous works such as $\uarch[pro]{\cdot}{}$ for ProSpeCT\cite{daniel2023prospect} or $\uarch[tt]{\cdot}{}$ for taint-tracking sandboxing schemes such as STT~\cite{yu2019speculative}. %
    }\label{figure:guarantees}
\end{figure}

\section{Restricting Attacker Observations}
\label{sec:attackerobs}
To achieve RMI, we first restrict the attacker's observations using microarchitectural isolation to satisfy \sw{shm}. We describe our baseline microarchitecture, and how it enforces strong isolation between security domains with shared memory disabled.
From there, we show how we have enabled shared memory while carefully constraining what an attacker can observe to (speculative) accesses to shared memory i.e. \sw{shm}.

\subsection{Strong Microarchitectural Isolation}
\label{sec:hardwareisolation}

\begin{figure}[t]
\centering
\includegraphics[width=\columnwidth]{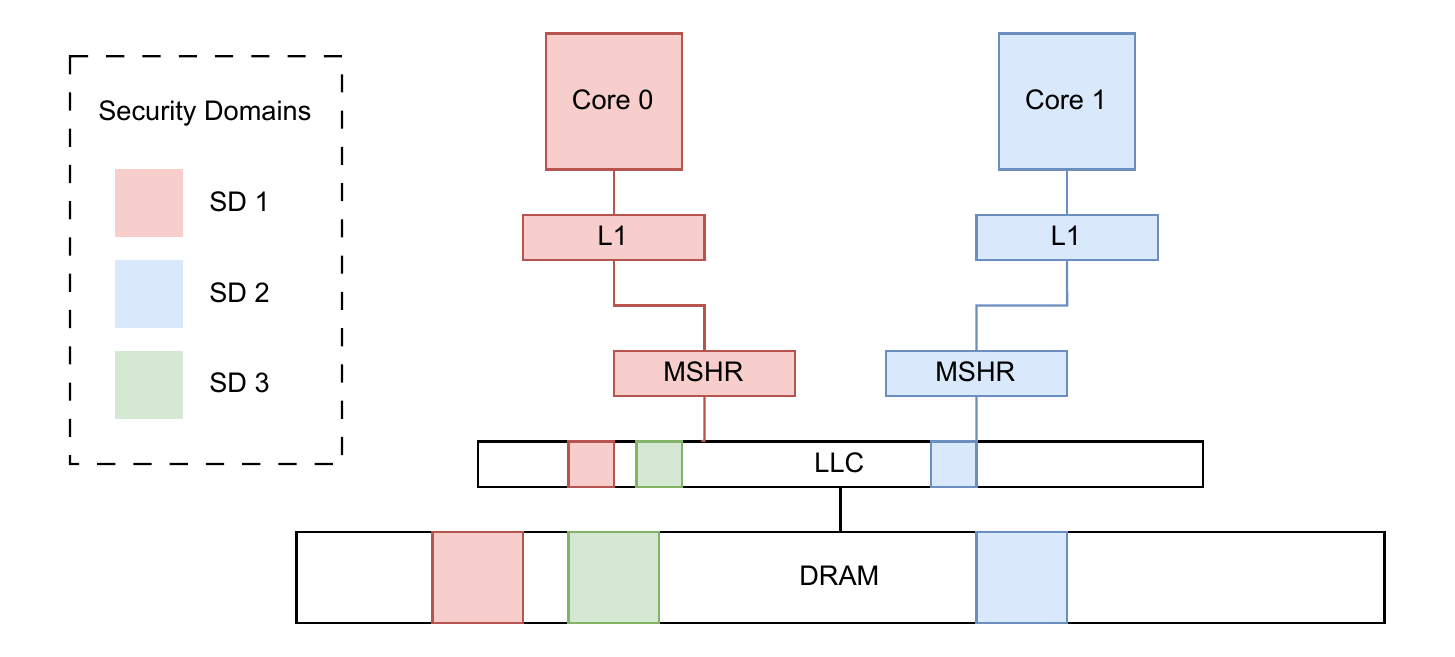}
\caption{Resource Partitioning Between Security Domains}
\label{fig:partitioning_all}
\centering
\includegraphics[width=\columnwidth]{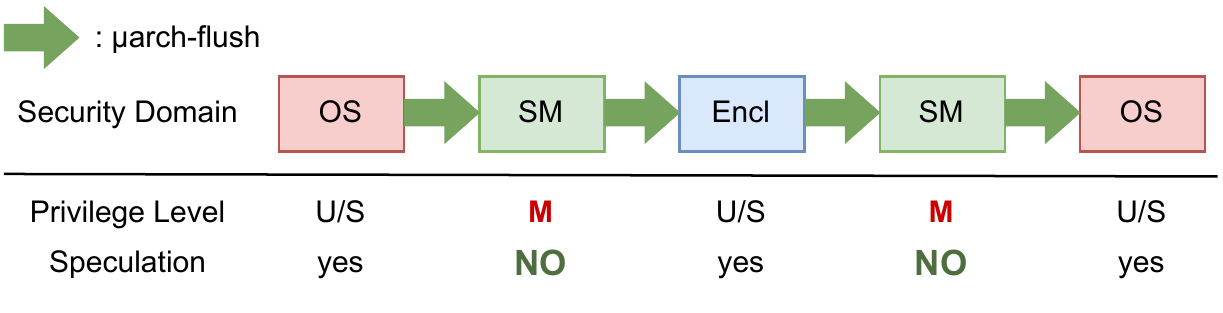}
\caption{Context Switching Between Security Domains}
\label{fig:context_switch}
\centering
\includegraphics[width=\columnwidth]{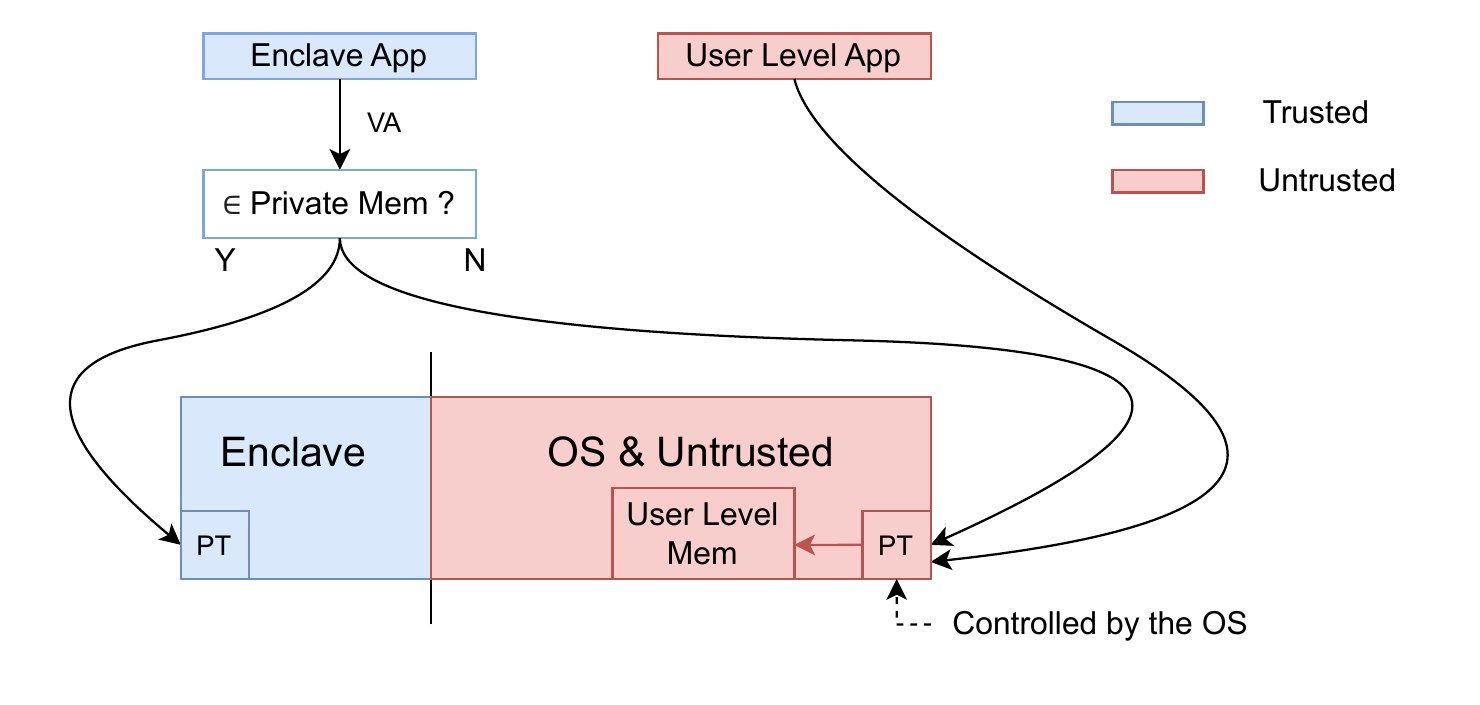}
\caption{The Dual Page Table Mechanism }
\label{fig:dual_pt}
\end{figure}

\noindent\textbf{Memory Isolation}
Physical Memory Protection (PMP) \cite{waterman2015risc} is the official RISC-V mechanism to partition physical memory and prevent unauthorized access between security domains (OS, enclaves, and SM).
However, it relies on accessing a series of registers, making it difficult to identify memory regions outside of the cores.
This makes it difficult to integrate features such as fine-grained LLC partitioning (see Section \ref{sec:dynamiccache}).
To allow for easy determination of a memory region from a physical address, 
we revive a mechanism from the
Sanctum paper \cite{costan2016sanctum}.
At its core, the mechanism divides memory into fixed-size physical memory regions (in our case 64 regions of 32MB) allocated by the SM to different security domains.
Each core uses two 64-bit registers as memory bitmaps, storing access rights for private and shared memory of the running security domain.
Citadel adopts this mechanism, enforcing checks on all memory accesses, including speculative and page-walk operations.
Figure \ref{fig:partitioning_all} illustrates this isolation, showing that only currently running security domains can access their corresponding DRAM regions.
\\

\noindent\textbf{Microarchitectural Isolation}
We base our design on MI6 \cite{bourgeat2019mi6}, which enforces strong microarchitectural isolation through partitioning, flushing, and disabling speculation in the SM.
First, all spatially shared microarchitecture is partitioned, including the LLC, the LLC pipeline, Miss Status Handling Registers (MSHR) and other resources as illustrated in Figure \ref{fig:partitioning_all}.
In particular, the LLC is statically partitioned into 64 regions to match the 64 DRAM memory regions -- we will improve this scheme in Section \ref{sec:dynamiccache}.
We pad DRAM request timing to a fixed duration that matches the slowest request. This slightly increases latency, but masks side channels from opened rows and contention in the DRAM controller.
We also pad requests coinciding with memory refresh (periodic and independent of any secrets) to a fixed duration.
We leverage interrupt logic to interpose the SM when context switching between security domains and enforce the flushing of temporally-shared microarchitecture such as all cores' pipeline state, L1 caches, TLBs, Translation Caches and MSHRs (see Figure \ref{fig:context_switch}).
The processor does not support hyper-threading and therefore is not vulnerable to related side channels such as port contention.
The processor does not include a prefetcher or SIMD units so we are not susceptible to Spectre variants exploiting them.
Finally, Citadel is not susceptible to Meltdown-style attacks, as the processor does not speculate across privilege levels (switching to a different privilege mode flushes the re-order buffer, making sure it never contains two instructions from different privilege levels).
The exhaustive list of side channels and defense mechanisms covered by MI6 can be found in \cite{bourgeat2019mi6}.

\noindent\textbf{Disabling Speculation in the SM}
The SM executes in machine mode (see Figure \ref{fig:context_switch}) which gives it access to all physical memory.
Following MI6 \cite{bourgeat2019mi6}, speculation in the SM is disabled to prevent TES attacks. 

\subsection{Enabling Shared Memory}
\label{sec:vmanddualpt}

Existing mechanisms like MMIO or non-cacheable memory~\cite{schwarz2020context, costan2016intel} are insufficient to prevent TES attacks.
They prevent some speculative memory accesses, but rely on physical addresses or page table entry bits, requiring speculative accesses to page table entries that leak address information.
Instead, we decide to adopt Sanctum's dual-page-table mechanism (Figure \ref{fig:dual_pt}) as it makes it possible to differentiate between private and shared memory based solely on virtual addresses.
Two new registers define the enclave's private virtual memory range.
When setting up the enclave, the SM ensures that physical memory mapped as private is owned by the corresponding security domain.
Memory access translation uses one of the two page tables:
\emph{Enclave page table}: For virtual addresses within the enclave memory range. The SM ensures that this table is stored in enclave private memory.
\emph{Shared page table}: For virtual addresses outside the private range (shared memory). This table is stored in shared memory.
\noindent During address translation, additional bound checks ensure that all memory addresses that are accessed belong to the same security domain (enclave or shared memory) as the translated address.
This mechanism can also be used to enable private shared memory between enclaves (not visible to the OS).
The SM simply needs to ensure that the root page table and shared memory region(s) are not accessible to the OS.
If shared memory with the OS is also required, some page tables can be located in the OS memory.

\subsection{Disentangling Private and Shared State}
\label{sec:disentangle}
We now need to carefully constrain what the attacker can observe on (speculative) accesses to shared memory.
As a design principle, read-only state used by both shared and private memory needs to be tagged to prevent aliasing while
core-private microarchitectural state entangled with adversary state through a coherence protocol need to be partitioned.
We highlight examples relevant to our design.

\noindent\textbf{Tagged ATC and TLB for Correct Address Translations} 
Both the Address Translation Cache (ATC) and Translation Lookaside Buffer (TLB) are susceptible to cross-domain aliasing.
The ATC, which caches intermediate page walker results, relies on a few upper virtual address bits, while TLB aliasing can occur with large pages if the private virtual memory range is smaller than the page granularity.
To prevent mistranslations, we tag ATC and TLB entries to indicate their association with private or shared memory.

\noindent\textbf{Speculative Transmitters in the L1-D Cache}
\label{sec:l1bypass}
While shared memory cannot be accessed speculatively, private memory can, potentially leading to speculative eviction of shared memory.
Unfortunately, the cache coherency protocol makes some private L1 microarchitectural state observable to an attacker, as writing to a memory location cached by another core takes longer than writing to an uncached location, effectively creating a side channel.
This vulnerability enables Spectre-style attacks through coherency in the L1-D cache.
In our case, partitioning the L1 would significantly impact overall system performance, including non-protected programs, due to the cache's latency sensitivity.
Instead, we opted to bypass the L1-D cache for enclave accesses to shared memory.
Because LLC is only reconfigured when no enclaves are running and the L1-D is flushed on context switch, we avoid issues related to cache coherency.
This decision is further justified by the limited temporal reuse of shared memory accesses and the observed impact on performance (see Section \ref{sec:evalsharedmemory}).
\\

\noindent\textbf{Security Analysis}
As we tagged read-only shared resources and the L1 is the only core-private state entangled with adversary state through a coherence protocol, we do not need to worry about other core-private structures.
Consequently, the attacker is only able to observe state of the shared LLC.
This effectively limits the attacker's observations to the trace of accesses, and addresses of accesses, to shared memory i.e.~\sw{shm}.
\section{Restricting Speculative Leakage}
\label{sec:controlledspec}
With attacker observations restricted to \sw{shm}, we can now focus on enforcing the execution model.
Observable traces when running a given program on our design should not leak more information than when running the same program non-speculatively i.e. using \sw{seq}.
This can be achieved through a combination of mechanisms for controlled speculation and program analysis.
We show two different approaches: preventing speculative shared-memory accesses altogether or ensuring that their addresses do not leak secrets.
We illustrate these approaches by building two minimal hardware mechanisms to enforce RMI for our enclave programs.
Safe mode achieves RMI without requiring any program analysis and Burst mode enables better performance on specific code snippets at the cost or requiring minimal code analysis.

\begin{figure}[t]
\centering
\includegraphics[width=\columnwidth]{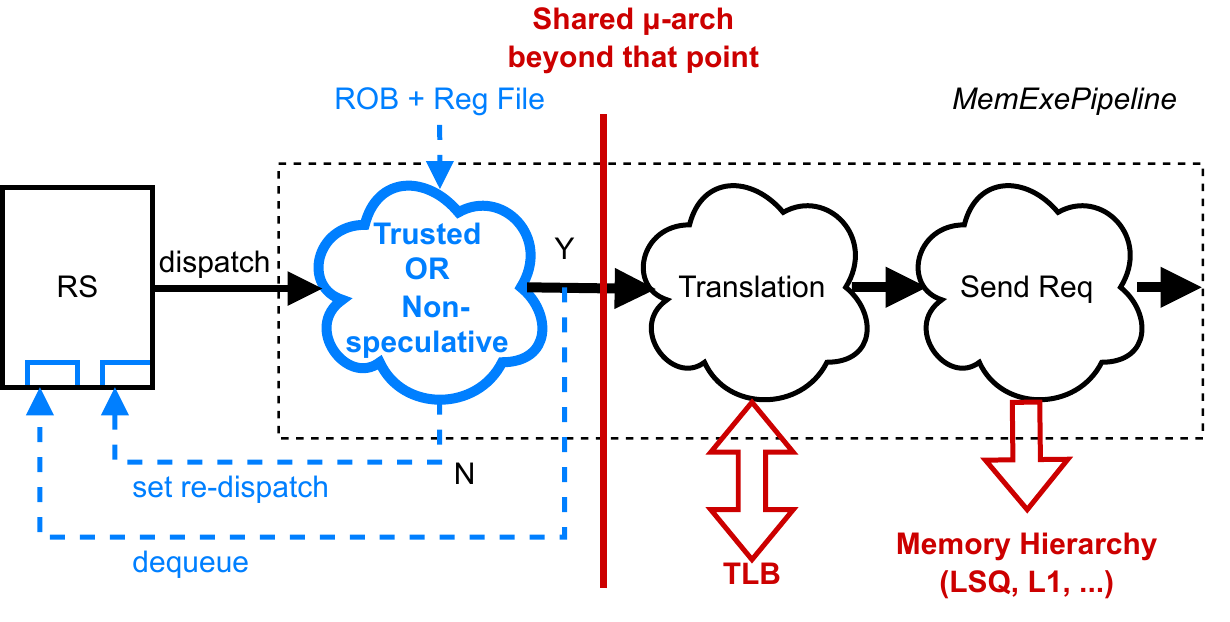}
\caption{Changes to the Memory Execution Pipeline in Citadel to enable Safe mode. New elements are shown in blue.}
\label{fig:MemExe}
\centering
\includegraphics[width=\columnwidth]{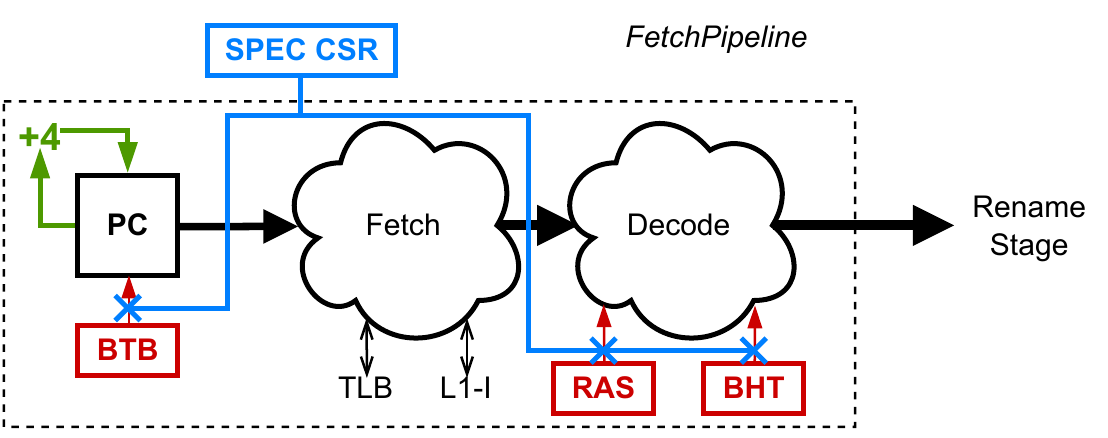}
\caption{Changes to the Fetch Pipeline in Citadel to enable Burst mode. New elements are shown in blue.}
\label{fig:fetchpipe}
\end{figure}

\subsection{Safe mode: Preventing Speculative Shared Memory}
\label{sec:mechanism1}
Our first approach consists of preventing speculative shared memory accesses altogether, effectively removing any speculative transmitters that can reach the attacker.
This approach enforces RMI for any program without extra program analysis, making it a great foundational mode to use by default.
We illustrate this principle by building a simple mechanism.
\emph{Safe} mode identifies unsafe shared memory accesses early in the memory execution pipeline, delaying their translation and processing until they become non-speculative (i.e., reach the head of the reorder buffer (ROB)).
This is similar \emph{in spirit} to delay-on-miss~\cite{sakalis2019efficient} but only delays \emph{shared} memory accesses and cannot leverage the L1 cache for reasons expressed in section \ref{sec:disentangle}.
A primer on out-of-order processors can be found in Appendix \ref{apendixooo} and microarchitectural details for Safe mode implementation can be found in Appendix \ref{app:safemode}.
Because the private memory range is defined in virtual memory, we can perform this check without any address translation, preventing information leakage through shared microarchitecture.
Safe instructions proceed to the next \MemExePipeline{} stage; unsafe ones are squashed and signaled to the reservation station so they are re-dispatched once they become non-speculative.
\\

\noindent\textbf{Security Analysis}
An attacker is only able to observe the trace of non-speculative accesses to shared memory. 
As a result, under our threat model, $\uarch[safe]{\cdot}{} \vdash \contract[shm]{\cdot}{seq}$ i.e. \textbf{Safe mode satisfies RMI}.

\subsection{Burst mode: Eliminating Information Leakage}
\label{sec:mechanism2}
Our second approach, called \emph{Burst} mode, is a hardware-software co-design that can improve performance for specific code snippets.
It allows some speculative accesses to shared memory (i.e. $\uarch[burst]{\cdot}{} \vdash \contract[shm]{\cdot}{stl}$) but ensures that it only runs programs that will not leak secrets (i.e. only runs $p$ if $p \vdash NI(\contract[shm]{\cdot}{seq} \Rightarrow \contract[shm]{\cdot}{stl})$).
This last step requires some static code analysis, a sometimes difficult problem (see Section \ref{sec:relatedwork}).
In order to simplify this analysis, we apply two simple design principles: 1) we ensure the (speculative) control flow of our code snippet is self-contained 2) we constrain speculation to a simple model where it is not attacker-controlled. 
The first principle indicates that we should deactivate any predictors that enable speculative jumps to arbitrary locations and guard our code snippets with speculative barriers.
The second stems from the fact that disabling more predictors helps simplify program analysis.
Following these principles, we implement a simple static branch predictor, which enables shared memory pipelining on Citadel (disabled in \emph{Safe mode}).
According to our first guidelines, we disable the branch target buffer (BTB), and the return address stack (RAS) (Figure \ref{fig:fetchpipe}).
This is done on demand by writing to a new CSR, using a write instruction that also doubles as a speculation barrier.
Following our second principle, we also disable the branch history table (BHT), effectively implementing straight-line speculation \cite{straightlinespecARM, guarnieri2021hardware} where the only prediction is the hardcoded ``\texttt{pc+4}'', predicting the next instruction without a change in control flow.

\subsection{Code Analysis for Burst mode}
Not all programs satisfy RMI when run in Burst mode. Take the following pathological code:
\noindent\begin{minipage}[]{0.25\textwidth}
\begin{lstlisting}[language={[RISC-V]Assembler}]
  jal far_away_function
  lw a0, 0(secret_address)
\end{lstlisting}
\end{minipage}
Here the secret address might be leaked under Burst mode while it would never be accessed during non-speculative execution.
To use Burst mode safely, we must perform code analysis to ensure snippets are: 1) self-contained to bound the analysis and 2) leak no secrets under straight-line speculation.
Self-containment is verified by checking that control flow remains confined between the instructions turning Burst mode on and off, i.e. no indirect branch and all branch destinations are valid and within the code snippet.
This constraint simplifies our analysis by limiting its scope to the code snippet.
We then verify that the code does not leak secrets under straight-line speculation.
Load, store, and branch instructions are considered potential speculative transmitters and the accessed register are marked as leaked.
For each transmitter, we perform a backward pass to determine dependencies of the leaked value and marking them as leaked.
This pass explores all possible earlier execution paths, including speculative and non-speculative paths when control flow deviates from pc+4.
Our simple speculation model prevents an explosion in possibilities, making this analysis tractable.
If a leaked value depends on another memory value, verification conservatively fails to prevent Spectre-like gadgets.
On each path, dependencies are \emph{declassified} as public if their values are leaked through a non-speculative load or store (we conservatively assume no declassification through branches).
We also prune paths that have been already explored if the set of leaked registers has not changed, and execution is not speculative, as we are not at risk of missing new leakage.
The backward pass yields a set of registers whose initial values could potentially be leaked, along with the corresponding speculative execution paths.
We can then identify conditions under which these paths are incorrect (i.e., not followed in non-speculative execution).
A code snippet fails verification if any non-public value can be leaked.
We developed a Python tool to automate this analysis.

\begin{figure}[htbp]
\begin{minipage}[]{0.25\textwidth}
\begin{lstlisting}[language={[RISC-V]Assembler},
  numbers=left,
  numberstyle=\tiny,
  numbersep=5pt]
  csrwi MSPEC, BURST_ON
  add	a2,a0,a2
  bgeu	a0,a2,.end # len!=0
.loop:
  lbu	a4,0(a1) # leaks src
  add	a1,a1,1
  add	a0,a0,1
  sb	a4,-1(a0) # leaks dest
  bne	a1,a2,.loop #leaks len
.end:
  csrwi MSPEC, BURST_OFF
\end{lstlisting}
\end{minipage}
\begin{minipage}[]{0.22\textwidth}
\begin{lstlisting}[language={[RISC-V]Assembler},
  numbers=left,
  numberstyle=\tiny,
  numbersep=5pt]
  add	a2,a0,a2
  bgeu	a0,a2,.end # len!=0
  csrwi MSPEC, BURST_ON
.loop:
  lbu	a4,0(a1)
  add	a1,a1,1
  add	a0,a0,1
  sb	a4,-1(a0)
  bne	a1,a2,.loop
  csrwi MSPEC, BURST_OFF
.end:
\end{lstlisting}
\end{minipage}
\caption{RISC-V Assembly code for 
\code{memcpy(a0=dest,a1=src,a2=len)}. Left hand-side does not satisfy RMI under Burst mode while the right-hand side does.}
\label{fig:memcpy-assembly}
\end{figure}

\noindent\textbf{\texttt{memcpy} example}
The default assembly can be seen on the left-hand side of Figure \ref{fig:memcpy-assembly}.
Here our tool identifies 3 potential transmitters line 5, 8 and 9.
Let's perform the backward pass for the branch on line 9 initially marking registers \code{a1} and \code{a2} as leaked.
We roll-back execution through line 8 to 6.
Line 6 overwrites \code{a1} which is declassified but replaced by its dependencies i.e. \code{a1}.
On line 5, a transmitter accesses \code{a1} but non-speculatively so it is not declassified.
At line 5, we encounter two possible paths: one from line 9 (which we prune as it is already explored), and another speculative path from line 3.
The analysis reveals a potential straight-line execution path from line 1 to 9 where the initial values of \code{a1} and \code{a2} (i.e., \code{src} and \code{len}) could be leaked.
This vulnerability occurs as Burst mode incorrectly speculates past the branch on line 3, which happens when \code{len==0}.
Our tool also reveals other leaks for \code{dest} and \code{src} under the same conditions.
We add a guard before entering Burst mode to check \code{len!=0} which generates safe assembly (right hand side).
The tool also detected self-containment issues with over-optimized compiler output (loop unrolling and rogue code block placements) where Burst mode would have been insecure.

\noindent\textbf{More Complex Analysis Tools}
The conditions we check on our code snippets are sufficient
but conservative.
To support more complex code patterns or speculation modes, more sophisticated tools like Kasper \cite{johannesmeyer2022kasper}, KleeSpectre \cite{wang2020kleespectre}, and Spectector \cite{guarnieri2020spectector} could potentially be adapted.
These tools usually do not scale well to large code bases and assume simplified speculation models.
These two limitations are circumvented by the design principles we presented.
\\

\subsection{Burst Mode Security Analysis}
\label{sec:burstsecanalysis}
The attacker observations are limited to \sw{shm}. 
Burst mode only allows for straight-line speculation, and control flow is self-contained.
That means that all reachable speculative paths are described by \sw{stl}. 
Under our threat model, $\uarch[burst]{\cdot}{} \vdash \contract[shm]{\cdot}{stl}$.
\\

\noindent\textbf{Extending Hardware Semantic With Static Software Analysis}
We extend hardware semantics to capture a static-software-analysis step.
Let's consider a static analysis tool \hw{tool} that takes any program $p$ as an input and returns \texttt{true} if the program passes the static analysis and \texttt{false} otherwise.
Given a static analysis program \hw{tool}, and a hardware semantic $\uarch[]{\cdot}{}$, for all program $p$, we define
\begin{align*}
    \uarch[]{p}{tool} = \begin{cases}
        \uarch[]{p}{} & \text{if } tool(p) = \texttt{true} \\
        \varnothing & \text{otherwise}
    \end{cases}
\end{align*}
Where $\varnothing$ represents the hardware semantic that only returns empty traces.
This can be easily implemented by adding a condition in the SM that verifies the (precomputed) result of \hw{tool}($p$) before starting the enclave.
An interesting special static analysis program is $\hw{\top}$ which returns true for any program.
\\

\noindent\textbf{Qualifying Our Static Analysis}
We call the static analysis used for Burst mode \hw{sta}.
Our tool assumes the \sw{shm} leakage model and only accepts programs that will not leak more secrets under straight-line speculation \sw{stl} than under sequential execution \sw{seq}.
That is, for a program $p$, \hw{sta} returns \texttt{true} if $p \vdash NI(\contract[shm]{\cdot}{seq} \Rightarrow \contract[shm]{\cdot}{stl})$, and \texttt{false} otherwise.
\\

\noindent
This gives us $\uarch[burst]{p}{sta} \vdash \contract[shm]{\cdot}{seq}$ (see proof in appendix~\ref{app:proofburstmode}).
In summary, \textbf{Burst mode satisfies RMI.}
\\

\noindent\textbf{Secure Composability}
Because our code snippets (speculative) control flows are self-contained, they can be safely reused and composed with code using other execution modes.
Developers can leverage pre-analyzed performance libraries (e.g., \texttt{memcpy}) without additional instrumentation or analysis.
In our design, we show that Burst mode on key code snippets significantly improves overall performance (see Section \ref{sec:evalsharedmemory}).
\section{Building and End-to-end Platform}

\noindent\textbf{Integrating a Reconfigurable LLC Partitioning Scheme}
\label{sec:dynamiccache}
We build a fine-grain set-partitioning scheme similar to \cite{townley2022composable,saileshwar2021bespoke} to support our 64 memory partitions.
Thanks to some of Citadel design choices, we are able to circumvent challenges in end-to-end integration ignored in previous work.
For instance, our memory region IDs are directly accessible through physical memory address (Section \ref{sec:hardwareisolation}) which makes it possible to support shared memory between cores without extending every structure or buses in the memory hierarchy to support security domain ID.
Additionally, we can leverage hardware software co-desing to ensure the LLC will not be put in an incoherent state during reconfiguration.
A request for reconfiguration always goes through the SM which will check 
that the mapping is correct and fits within the LLC.
The SM will check that no enclaves are running and preempt other cores to ensure no requests will be sent to the affected memory regions during the operation.
Implementation details can be found in Appendix \ref{app:dynamiccache}.
To our knowledge, we are the first to implement such a scheme in RTL and to integrate it in a system.
\\

\noindent\textbf{Fine-grained LLC Flushing}
We also implement a hardware-software fine-grain flushing mechanism.
We add a simple piece of logic connected to the cache hierarchy at the same level than the DRAM, which always sends back zeroes.
It is mapped to a physical address space the size of the DRAM making it possible to easily build and access eviction sets in software.
This means that flushing a small LLC partition is much faster than flushing a big one.
Implementation details can be found in Appendix \ref{app:llcflush}.
\\

\noindent\textbf{Building the Software Infrastructure}
Our software architecture, illustrated early in the paper in Figure \ref{fig:computing_stack}, comprises several key components. 
We implement the SM at RISC-V's highest privilege level using assembler and C, alongside a \emph{Mini-SM} - a downsized, enclave-private copy of the global SM.
A Linux Kernel module enables user programs to request SM calls and set up enclaves, with Linux managing resources.
Additionally, we develop a Secure Bootloader and an end-to-end attestation mechanism, enabling unique enclave identification, shared secret establishment, and encrypted private data exchange.
Implementation details regarding our entire software infrastructure can be found in Appendix \ref{app:softwareinfrastructure}.

\begin{table}
\begin{tabular}{ | m{5em} | m{17em} | } 
  \hline
  Frontend & 2-way superscalar, 256 entries BTB, 8 entries RAS, tournament predictor \\
  \hline
  ROB &  64 entries, 2-way insert/commit \\
  \hline
  Reservation Stations & 2 stations for ALU (16 entries), 1 for FPU (16 entry), 1 for memory (16 entries) \\
  \hline
  L1 (I/D) &  32KB, 8 ways, 8 outstanding requests \\
  \hline
  Ld-St Unit & 24 LdQ entries, 14 StQ entries, 4 SB entries\\
  \hline
  LLC (L2) & 1MB, 10 cycles latency, 16 ways, max 16 outstanding req. \\
  \hline
  Memory &  120 cycles latency (24 outstanding reqs) \\
  \hline
\end{tabular}
    \caption{Citadel's microarchitectural configuration.}
    \label{tab:mi6param}
\end{table}
\begin{table}
    \centering
    \begin{tabular}{|p{3cm}|r|r|}
        \hline
         Component & LOC & Size\\
         \hline\hline
         \rowcolor{LightRedbis}
         Processor\cite{bourgeat2019mi6} (+mod) & $\sim60$ KLOC (+ 350) & N/A \\
         \hline
         \rowcolor{myblue}
         Bootloader & 5236 & 55KB\\
         \hline
         \rowcolor{myblue}
         Security Monitor & 9287 & 97KB\\
         \hline
         \rowcolor{LightRedbis}
         Mini-SM & 2301 & 5.6KB\\
         \hline
         \rowcolor{LightRed}
         Linux & $\sim20$M & 157MB\\
         \hline
         \rowcolor{LightRed}
         SM Kernel Mod. & 507 & 505KB\\
         \hline
         \hline
         \rowcolor{myblue}
         TCB Software & 14523 & 152KB \\
         \hline
    \end{tabular}
    \caption{Software Component Size and TCB Breakdown. Mini-SM is a subset of the SM code. 
    Lines in red correspond to elements excluded from the TCB.}
    \label{tab:tcbloc}
    \smallskip
    \centering
    \small
    \begin{tabular}{|p{1.2cm}|r|r|}
        \hline
         Modules & LUTs & FFs\\
         \hline\hline
         Top & 410K (\textcolor{blue}{+3.0\%} | \textcolor{orange}{+1.8\%}) & 251K (\textcolor{blue}{+4.4\%} | \textcolor{orange}{+2.6\%})\\
         \hline
         Core & 169K (\textcolor{blue}{+1.6\%} | \textcolor{orange}{+0.3\%}) & 91K (\textcolor{blue}{+3.4\%} | \textcolor{orange}{+1.0\%})\\
         \hline
    \end{tabular}
    \caption{Hardware overhead for the top module and the core, lookup tables (LUTs) and flip flops (FFs). First comparison is to Riscy-OOO while second is to MI6 with the parameters of Table \ref{tab:mi6param}.}
    \label{tab:hwoverhead}
\end{table}

\section{Evaluation}
\label{sec:eval}

\noindent\textbf{Experimental Setup}
Citadel's processor is a modified out-of-order, superscalar, 2-core based on Riscy-OOO~\cite{zhang2018composable} and MI6~\cite{bourgeat2019mi6}.
The parameters used for the cores and the memory subsystem are given in Figure~\ref{tab:mi6param}, and kept small to keep the synthesis and simulation time small.
Citadel runs on an AWS EC2 F1 instance, equipped with a Virtex UltraScale+ FPGA. 
On this FPGA board, Citadel can be clocked at $30$ MHz, which is similar to the baseline design.
We use the RISC-V tool chain with GCC 12.2.0.
We compile all software with the -O3 flag. 
We run Linux Kernel 6.2.0.
All numbers are collected by running the benchmarks on the FPGA using performance counters on an average of ten runs. 
\\

\noindent\textbf{Implementation Data, TCB and Surface Overhead}
Number of lines of code (LOC) and binary size of Citadel's components are presented in Table \ref{tab:tcbloc}.
Due to their limited functions, the SM and the Bootloader can be kept small and the total software Trusted Code Base (TCB) is less than 15K LOC.
This helps minimize potential bugs and reduce the software attack surface.
It can also help make future formal verification efforts tractable.
The final TCB size needs to include the enclave's binary, which depends on the application.
Table \ref{tab:hwoverhead} shows the area overhead for Citadel compared to Riscy-OOO\cite{zhang2018composable} and  MI6\cite{bourgeat2019mi6}.
Most of the overhead comes from enforcing microarchitectural isolation (i.e. MI6) or from the modifications in the LLC with, compared to Mi6, the addition of an arbiter, a fine-grained partitioning mechanism and the L1 bypass mechanism.
The overhead for the core compared to MI6 is minimal (+0.3\% LUTs and +1.0\% FFs), which gives us an upper bound for Safe mode and Burst mode.
This is significantly lower than other mechanisms for secure speculation (see Section \ref{sec:relatedwork}).

\begin{table}[ht]
    \centering
    \begin{tabular}{|c|c|c|}
    \cline{2-3}
    \multicolumn{1}{c|}{} & \textbf{Memcpy} & \textbf{Random} \\ \hline
    \textbf{Baseline} & 343928  &  273164 \\ \hline
    \textbf{Safe} & 595362 (+73\%) & 1447543 (+430\%) \\ \hline
    \rowcolor{mygreen!35}
    \textbf{Safe+Burst} & 343291 (+0\%) & 270394 (-1\%) \\ \hline
    \textcolor{red}{\textbf{Safe+Burst+L1}} & 344163 (+0\%) & 277969 (+2\%) \\ \hline
    \end{tabular}
    \caption{Microbenchmarks for \texttt{memcpy} and random accesses to shared memory. \texttt{memcpy} copies 128MB from shared memory to private memory. Baseline is \texttt{memcpy} from an to private memory with speculation enabled. \emph{Random} accesses 10240 elements in a 128MB memory region.
    We experiment with \emph{insecurely} enabling the L1 for shared memory accesses.}
    \label{tab:resultmemcpy}
\newcolumntype{C}[1]{>{\centering\arraybackslash}m{#1}}
     \begin{NiceTabular}{|C{2.7cm}|C{1.2cm}|C{1.2cm}|C{1.4cm}|}[hvlines, corners=NW]
        \CodeBefore
        \rowcolor{gray!35}{2}
        \rowcolor{mygreen!35}{4}
        \Body
            \multicolumn{1}{c|}{} & \textbf{ECC} & \textbf{ML} & \textbf{ML EXT}\\
            \Block{}{\textcolor{red}{\textbf{{Baseline}}}} & 98599278 & 406765555 & 406765555 \\
            \Block{}{\textbf{Safe Only}} & 102433204 (+4\%) & 419853901 (+3.2\%) & 1119486865 (+175\%)  \\
            \Block{}{\textbf{Citadel\\(Safe+Burst)}} & 100411982 (+1.8\%) & 421332976 (+3.5\%) & 426527197 (+4.8\%) \\
    \end{NiceTabular}
    \caption{Performance overheads for applications that access shared memory. Our elliptic curve cryptography (ECC) benchmark signs a total of 1MB of data split in 3000 packets. Our ML benchmark evaluate private inference over 1024 encrypted requests. ML EXT is the same but the public model weights are in shared memory.
    Requests are sent by an untrusted app using message passing through shared memory. 
    Insecure Baseline is the same application directly linked into the untrusted app.
    }
    \label{tab:resultcrypto}
\end{table}

\subsection{Microbenchmarks}
\label{sec:evalsharedmemory}

We evaluate two \emph{pathological} microbenchmarks with distinct but shared memory access patterns: \emph{Memcpy} (sequential access, copying 128KB from shared to enclave private memory) and \emph{Random} (10,240 random accesses within a 128KB region).
Results can be found in Table \ref{tab:resultmemcpy}.
We compare these with a baseline of private memory accesses.
For all tests, we disable LLC partitioning and use 4KB page mapping.
Safe mode alone significantly increases overhead (+73\% for Memcpy, +430\% for Random) due to disabling the pipelining of shared memory accesses. 
Activating Burst mode for both access types (after verifying security with our code analysis tool) dramatically reduces this overhead, approaching baseline performance.
Note that verification is a one-time task and that Burst-enable \emph{memcopy} can now be used by other code \emph{without further program analysis}.
We also experiment with (insecurely) disabling the L1 bypass for shared memory accesses, which, for these benchmarks, has negligible performance impact. 
Indeed, for \texttt{memcpy}, adjacent pipelined loads will always be resolved on the next cycle (in the L1 or the LLC) and similarly than for Random, data is not re-accessed.
For Random, these results show that polluting the L1-D with random memory accesses slightly degrades performance.

\subsection{Evaluating the End-to-End Platform}

We evaluate the platform usability by porting existing applications, showing how Burst mode can be easily leveraged to gain performance at the cost of minimal code annotation and analysis.
Our shared memory mechanisms only incur performance overheads when the enclave program accesses shared memory, making evaluation mostly relevant for programs with private and shared memory (e.i. not SPEC).
Note that for non-enclave programs, the only overhead is due to the LLC arbiter and is minimal on our two-cores prototype. We measure $\approx$0.1\% overhead compared to Riscy-OOO~\cite{zhang2018composable} on our different benchmarks when run as insecure baselines. 
\\

\noindent\textbf{Securing a Crypto Library (ECC)}
\label{sec:evalcrypto}
We write a wrapper around an ED25519 RISC-V library \cite{ed25519github}.
We add a queue to shared memory, so untrusted applications can send requests to the library to generate keys (that will stay in enclave memory), or to sign messages using previously generated keys.
In total, we need to implement 334 LOC of C, which is to be compared to the 4,514 LOC of the crypto library (7\% overhead).
With the enclave's code added, the total TCB is now 19,231 LOC.
The final crypto-enclave binary is 71KB.
As shown in Table \ref{tab:resultcrypto}, we see minimal overhead mainly due to message passing and copying inputs inside the enclave.
We instrument the SHA256 macro that performs the input copy to leverage Burst mode.
We successfully verify that the code snippet is secure using our analysis tool we reduce the overhead to 1.8\%.
For completeness, and because the ECC benchmark is implemented using constant-time programming \cite{reparaz2017dude, almeida2016verifying}, we also evaluate the performance overhead of disabling speculation completely in order to defend against "execution time" attacks (see Section \ref{sec:execution_time_SC}). We observe a +882\% overhead, orders of magnitude above our other mechanisms, which clearly highlights the trade-off Citadel offers between security and performance.
\\

\noindent\textbf{Private Inference (ML and ML EXT)}
\label{sec:evalprivateinference}
We convert a handwritten digit recognition model (trained on MNIST) to standard C code \cite{onnx2cgithub} and run private inference inside our enclave.
A user uses our remote attestation mechanism to verify the enclave identity and establish a secure communication channel.
When receiving requests through shared memory, the enclave decrypts the ciphertext, performs the inference task on the private input and sends back an encrypted output.
We also experiment with placing the model weights in shared memory (ML EXT).
This scenario is especially relevant as some models might be too big to fit in DRAM and require to leverage demand paging. 
Note that this type of access patterns would be hard to reproduce using message-passing primitives (e.g. DMA accesses).
In this scenario, the integrity of the model weights is not protected.
In total, we need to implement 227 lines of straightforward C, which is to be compared to the 1536 LOC of the neural network, the 410 LOC of the AES library and the 4386 LOC of ED25519 we also include
for encryption and key exchange.
The TCB is now 21K LOC and the final enclave binary measures 54KB.
Setting up the enclave, represents a (one-time) cost similar to performing 235 inferences.
For a given request, runtime is split 55\% and 45\% between AES and inference.
Table \ref{tab:resultcrypto} show that overheads are reasonable but can get up to +175\% when placing the model in shared memory.
We use our Burst-instrumented \texttt{memcpy} function but see no performance improvement as only small messages are copies.
Nevertheless, on ML EXT, we instrument the inference code to leverage Burst (+2 LOC).
Our code analysis tool revealed a Spectre-like gadget as the pointer to the tensor is first dereferenced before the tensor values are accessed.
We made sure to deference the first pointer before turning on Burst mode (+1 LOC), enough to pass program analysis.
We observe performance overhead drops down to 4.8\%. 
These results highlight the versatility of shared memory and the tradeoff between performance and requiring program analysis.
Finally, insecurely disabling L1-bypass for shared memory on ML EXT only show modest performance gains (+4.3\% down from 4.8\% ).
\\ 

\noindent\textbf{MicroPython Runtime}
\label{sec:evalmicropython}
We port MicroPython \cite{microphython} inside an enclave
which serves as a small benchmark to demonstrate the expressivity and flexibility of the platform.
We redirect the console to shared memory and a hash of the standard output is extended on each character print.
The hash of the console transcript is signed on the enclave exit using ECC and the enclave's own private keys (obtained through the local attestation mechanism).
In total, we modify 128 LOC of MicroPython.
With the enclave's code added, the TCB is now 65K LOC, most of it (70\%) being MicroPython.
We successfully run our attested runtime on a simple primality test written in Python.

\subsection{Evaluating LLC Dynamic Partitioning}
\label{sec:evalllcpartitioning}

We evaluate dynamically partitioning the LLC on our cryptographic enclave benchmark.
We compare 1)
an insecure baseline where the LLC is not partitioned, 2) a secure configuration where the LLC is uniformly statically partitioned (MI6 scheme) and 3) a non-uniform partitioning set up dynamically using our scheme.
For the non-uniform configuration, we split the 1MB of LLC as follows: we give 256KB to the enclave, 16KB to the SM, 128KB to the OS, 128KB to the region used for shared memory, 128KB to the region storing page tables, and 1KB each to every other region.
We observe a 67.8\% overhead for the uniform partitioning and 0.08\% overhead for the non-uniform partitioning.
This is especially important as untrusted applications also benefit from it.
Modifying LLC partitioning (infrequent) takes on average 500K cycles.
Among these, 98\% of the cycles are spent flushing the LLC.
Resizing fewer regions is faster: we measure an average of 93 cycles to flush a single LLC set containing 16 ways.

\begin{table*}[t]
    \centering
    \footnotesize
    \setlength{\tabcolsep}{2pt}  
    \begin{minipage}[t]{0.48\textwidth}
    \centering
    \begin{tabular}{|>{\centering\arraybackslash}m{2.3cm}|*{4}{c|}c|c|c|c|}
    \hline
    \multirow{2}{*}[-5.5ex]{\begin{tabular}[c]{@{}c@{}}\textbf{Enclave}\\\textbf{Platforms}\end{tabular}} & 
    \multicolumn{4}{c|}{\textbf{Security}} & \multicolumn{1}{c|}{\textbf{Usability}} & \multicolumn{3}{c|}{\textbf{Implementation}} \\
    \cline{2-9}
    & \rotatebox{90}{\parbox{1.5cm}{\centering SC LLC}} &
    \rotatebox{90}{\parbox{1.5cm}{\centering SC Others}} &
    \rotatebox{90}{\parbox{1.5cm}{\centering TES}} &
    \rotatebox{90}{\parbox{1.5cm}{\centering Exec-Time}} &
    \rotatebox{90}{\parbox{1.5cm}{\centering Shared Memory}} &
    \rotatebox{90}{\parbox{1.5cm}{\centering HW}} &
    \rotatebox{90}{\parbox{1.5cm}{\centering SW. Infra.}} &
    \rotatebox{90}{\parbox{1.5cm}{\centering Apps}} \\
    \hline
    Commercial \cite{mckeen2013sgx, costan2016intel,cheng2024intel, sev2020strengthening,trustzone04, pinto2019demystifying, li2022design}& \red{V} & \red{V} & \red{V} & \red{V} & \green{Yes} & \green{TO} & \green{Yes} & \green{Yes} \\
    \hline
    Academic \cite{azab2011sice, brasser2019sanctuary, deng2014equalvisor, feng2021scalable, van2023cheri, ferraiuolo2017komodo, weiser2019timber} & \red{V} & \red{V} & \red{V} & \red{V} & \green{Yes} & \orange{R/S} & \orange{Some} & \orange{Some} \\
    \hline
    Sanctum \cite{costan2016sanctum} & \green{P} & \red{V} & \red{V} & \red{V} & \green{Yes} & \red{Sim} & \red{No} & \red{No} \\
    \hline
    KeyStone \cite{lee2020keystone}, CURE\cite{bahmani2021cure}, Hybrid\cite{enclavecaches} & \green{P} & \red{V} & \red{V} & \red{V} & \green{Yes} & \green{RTL} & \green{Yes} & \green{Yes} \\
    \hline
    MI6 \cite{bourgeat2019mi6} & \green{P} & \green{P} & \green{P} & \red{V} & \red{No} & \green{RTL*} & \red{No} & \red{No} \\
    \noalign{\hrule height 1pt}
    \textbf{Citadel} & \textbf{\green{P}} & \textbf{\green{P}} & \textbf{\green{P}}  & \textbf{\red{V}} & \textbf{\green{Yes}} & \textbf{\green{RTL}} & \textbf{\green{Yes}} & \textbf{\green{Yes}} \\
    \noalign{\hrule height 1pt}
    \end{tabular}
    \caption{Comparison with Other Enclave Platforms.
    SC: Side Channel, TES: Transient Execution SC, Exec-Time: Execution-Time SC,
    P: Protected, V: Vulnerable, TO: Taped-out, Sim: Simulator, R/S: RTL or Sim, RTL*: Incomplete}
    \label{tab:enclave-platforms}
    \end{minipage}%
    \hfill
    \begin{minipage}[t]{0.48\textwidth}
    \centering
    \begin{tabular}{|>{\centering\arraybackslash}m{2.1cm}|*{4}{c|}*{2}{c|}c|c|}
    \hline
    \multirow{2}{*}[-5.5ex]{\begin{tabular}[c]{@{}c@{}}\textbf{Defense}\\\textbf{Mechanism}\end{tabular}} & 
    \multicolumn{4}{c|}{\textbf{Security}} & \multicolumn{2}{c|}{\textbf{Usability}} & \multicolumn{2}{c|}{\textbf{Overhead}} \\
    \cline{2-9}
    & \rotatebox{90}{\parbox{1.5cm}{\centering TES Cache}} &
    \rotatebox{90}{\parbox{1.5cm}{\centering TES Others}} &
    \rotatebox{90}{\parbox{1.5cm}{\centering Register Values}} &
    \rotatebox{90}{\parbox{1.5cm}{\centering TES Exec-Time}} &
    \rotatebox{90}{\parbox{1.5cm}{\centering Annotation Secrets}} &
    \rotatebox{90}{\parbox{1.5cm}{\centering Code Analysis}} &
    \rotatebox{90}{\parbox{1.5cm}{\centering Hardware}} &
    \rotatebox{90}{\parbox{1.5cm}{\centering Performance}} \\
    \hline
    ProSpeCT \cite{daniel2023prospect} & \green{P} & \green{P} & \green{P} & \green{P} & \red{Yes} & \red{CT} & \red{High} & 0-45\% \\
    \hline
    SPT \cite{choudhary2021speculative} & \green{P} & \green{P} & \green{P} & \green{P} & \green{No} & \red{CT} & \red{Sim} & 45\% \\
    \hline
    STT \cite{yu2019speculative,jauch2023secure} & \green{P} & \green{P} & \red{V} & \green{P} & \green{No} & \green{No} & \red{High} & 46\% \\
    \hline
    SpecShield \cite{barber2019specshield} & \green{P} & \green{P} & \red{V} & \green{P} & \green{No} & \green{No} & \red{Sim} & 21-73\% \\
    \hline
    NDA \cite{weisse2019nda} & \green{P} & \green{P} & \orange{S} & \green{P} & \green{No} & \green{No} & \red{Sim} & 100\% \\
    \hline
    SpectreGuard \cite{fustos2019spectreguard} & \green{P} & \green{P} & \green{P} & \green{P} & \red{Yes} & \green{No} & \red{Sim} & 8-20\% \\
    \hline
    ConTExT \cite{schwarz2020context} & \green{P} & \green{P} & \green{P} & \red{V} & \red{Yes} & \green{No} & \red{Sim} & 0-338\% \\
    \hline
    InvisiSpec \cite{yan2018invisispec} + \cite{khasawneh2019safespec, sakalis2019efficient, li2019conditional, ainsworth2020muontrap, saileshwar2019cleanupspec} & \green{P} & \red{V} & \green{P} & \red{V} & \green{No} & \green{No} & \red{Sim} & 21-72\% \\
    \hline
    DOLMA \cite{loughlin2021dolma} & \green{P} & \green{P} & \green{P} & \green{P} & \green{No} & \green{No} & \red{Sim} & 42.2\% \\
    \hline
    Sequential & \green{P} & \green{P} & \green{P} & \green{P} & \green{No} & \green{No} & \green{Negl} & 882\% \\
    \noalign{\hrule height 1pt}
    \textbf{Citadel} & \textbf{\green{P}} & \textbf{\green{P}} & \textbf{\green{P}} & \textbf{\red{V}} & \textbf{\green{No}} & \textbf{\orange{Min}} & \textbf{\green{Small}} & \textbf{\green{0\% - 5\%}} \\
    \noalign{\hrule height 1pt}
    \end{tabular}
    \caption{Comparison with Hardware Defense Mechanisms for Transient Execution Side Channels (TES), P: Protected, V: Vulnerable, S: Some, CT: Constant Time, Sim: Simulator}
     
    \label{tab:defense-mechanisms}
    \end{minipage}
    
\end{table*}

\section{Related Work}
\label{sec:relatedwork}
\noindent\textbf{Security of Enclave Platforms}
\label{sec:enclaveplatforms}
Table \ref{tab:enclave-platforms} illustrates the landscape of existing enclave platforms. 
All commercial platforms~\cite{mckeen2013sgx, costan2016intel,cheng2024intel, sev2020strengthening,trustzone04, pinto2019demystifying, li2022design} have proven vulnerable to TES attacks~\cite{chen2019sgxpectre,van2020sgaxe,gotzfried2017cache,brasser2017software,van2018foreshadow,van2020lvi,weisse2018foreshadow,ryan2019hardware,xu2024trustzonetunnel}.
Despite this, most commercial platforms and academic works on TEEs~\cite{azab2011sice, brasser2019sanctuary, deng2014equalvisor, feng2021scalable, van2023cheri, ferraiuolo2017komodo, weiser2019timber, van2023cheri, li2021confidential, lee2022cerberus, park2020nested, steinegger2021servas} do not include these attacks in their threat models, shifting the burden to developers and deferring defense to software analysis tools (see lower).
Although some academic proposals such as Sanctum \cite{costan2016sanctum}, KeyStone \cite{lee2020keystone}, and Cure \cite{bahmani2021cure} incorporate some defenses against cache side channels, only MI6 \cite{bourgeat2019mi6} formally addresses TES attacks but disables shared memory, failing to address \prob{} comprehensively.
Citadel builds on the open source code provided in MI6 \cite{bourgeat2019mi6} and reuses many of its microarchitectural isolation mechanisms.
The MI6 paper focused on evaluating the \emph{performance} of these mechanisms on an FPGA
by running SPEC benchmarks with no security guarantees by using Linux system calls as hooks to project the cost of a potential context switch to the SM.
Citadel's performance numbers for SPEC would be similar to MI6's evaluation.
Citadel stands out as the only platform that offers protection against TES attacks while maintaining shared memory usability and providing a complete implementation with hardware, software infrastructure, and applications.
\\

\noindent\textbf{Software Analysis Tools}
\label{sec:softwaremitigations}
Commercial TEEs provide speculation barriers and other flushing instructions for manual placement in software~~\cite{TrustDomainIntel, ArmSpecualtiveVulnerabilities, AMDsoftwarespecualtion}.
However, these countermeasures~\cite{mosier2024serberus, cauligi2022sok, disselkoen2020finding, guanciale2020inspectre} incur significant overhead, might not be sufficient \cite{barberis2022branch}, and their reliable automated application remains an open challenge~\cite{LLVMdevSpeculativeLoadHardening,IntelAnalysisSideChannel, MicrosoftMSVC,wang2019oo7}.

Other tools aim to target more precise speculative security policies~\cite{guarnieri2021hardware,cauligi2022sok}, but mostly target constant-time crypto~\cite{barthe2021high, cauligi2020constant, daniel2021hunting, guanciale2020inspectre, guarnieri2020spectector, patrignani2021exorcising, vassena2021automatically} or the sandboxing scenario~\cite{cheang2019formal, narayan2021swivel, qi2021spectaint, shen2018restricting, jenkins2020ghostbusting, oleksenko2020specfuzz, kirzner2021analysis} which makes them difficult to adapt to address \prob{} (see Section \ref{sec:beyondsandboxing}).
It is also unclear how pieces of code analyzed with these tools and satisfying \emph{stronger} speculative properties than RMI should be composed, or how they should communicate with each other.
In fact, introducing any universal read gadget in the code base would break security, even for the carefully analyzed speculative constant-time~\cite{cauligi2020constant} cryptographic library.
Citadel presents one strategy to compose different speculative policies.
It can even compose programs that guarantee the ``pre-Spectre'' non-interference properties~\cite{barthe2014system}, without worrying about speculation, as we can guarantee that predictors are independent of any secrets through isolation.
Finally, analysis tools struggle to scale to entire codebases, especially when trying to prove complex properties, further complicating this composition problem~\cite{guarnieri2020spectector,wang2019oo7,johannesmeyer2022kasper,wang2020kleespectre}.
Citadel addresses these limitations by isolating execution modes, enabling secure shared memory, and limiting software analysis to small code snippets.

Other tools like Revizor \cite{oleksenko2022revizor, garcia2023fuzzing} make it possible to fuzz CPUs for speculative leaks. 
To our knowledge, current available implementations only support X86. 
Adapting Revizor to RISC-V and RMI is left for future work.

\subsection{Hardware Mitigations for TES attacks}
\label{sec:exisitngTEAmech}
Numerous academic proposals \cite{ainsworth2020muontrap, cheang2019formal, canella2019systematic, barber2019specshield, fustos2019spectreguard, yan2018invisispec, yu2019speculative, saileshwar2019cleanupspec, weisse2019nda, khasawneh2019safespec, guarnieri2020spectector, guarnieri2021hardware, mosier2022axiomatic,ainsworth2021ghostminion, daniel2023prospect, choudhary2021speculative, kiriansky2018dawg, sakalis2019efficient, li2019conditional, narayan2023going} leverage hardware mechanisms to counter TES attacks.
Why not use some of these mechanisms as primitive and combine them with, say, MI6~\cite{bourgeat2019mi6} to address \prob{}?
As shown in Table \ref{tab:defense-mechanisms}, each mechanism offers a different trade-off between security, usability, and performance.
Within that design space, Citadel and its simple mechanisms can efficiently address \prob{}.

As explained in Section~\ref{sec:beyondsandboxing}, mechanisms that focus on the \emph{constant-time}~\cite{daniel2023prospect, choudhary2021speculative} and \emph{sandboxing}~\cite{yu2019speculative,jauch2023secure,barber2019specshield, yu2020speculative} scenarios are not the best building blocks to address \prob{} efficiently.
Similarly, NDA\cite{weisse2019nda} ignores Specter-BTB and does not prevent leakage of
register secrets through single transient micro-op.
Other Hardware Mechanisms such as Context\cite{schwarz2020context}, OISA\cite{yu2018data} and SpecterGuard\cite{fustos2019spectreguard} require extensive manual and error-prone secret annotations, on which \emph{security} depends.
In contrast, Citadel' code annotation is only required for \emph{performance} and any manual error would be caught by our static analysis tool.
Many other mechanisms could be a good fit to be used as primitives to build a platform such as Citadel.
Invisible speculation schemes such as InvisiSpec~\cite{yan2018invisispec} and others~\cite{ainsworth2020muontrap, khasawneh2019safespec, kiriansky2018dawg, li2019conditional, saileshwar2019cleanupspec, sakalis2019efficient, ainsworth2021ghostminion} only address cache side channels.
Similarly, DOLMA\cite{loughlin2021dolma} could be used to address \prob{} and even provide guarantees against the execution time SC (see Section~\ref{sec:execution_time_SC}), \emph{if} augmented with micro-architectural isolation similar to MI6.
\\

\noindent\textbf{Hardware and Performance Overhead}
However, even when considered as primitives, existing hardware mechanisms are \emph{complex} and incur significant \emph{overheads}.
Few of these mechanisms have been implemented in hardware (FPGAs or silicon), and many attacks that break these defenses \cite{yang2023pensieve} exploit details that are often overlooked in simulation, such as faithful memory hierarchy representation.
The two mechanisms that have been implemented incur significant hardware overhead.
STT on BOOM \cite{jauch2023secure} adds +15.7\% LUTs and 6.2\% FFs, while ProSpeCT \cite{daniel2023prospect} adds +17\% LUTs and +6\% FFs, an overhead potentially prohibitive for commercial deployment, especially given that they would require extra isolation mechanisms to address the TEE threat model (i.e., privilege attacker).
In that sense, these overheads should be compared with the minimal surface overhead of implementing Safe mode and Burst mode on top of MI6: +0.3\% LUTs and +1.0\% FFs.

These complex mechanisms also incur significant performance overhead.
DOLMA presents an overhead of +42.2\% on average, while NDA + 100\% and STT + 46\% for their most conservative threat models (but closest to our).
In contrast, Citadel shows a modest 0-5\% overhead on our benchmarks.
\section{Conclusion}

We define \emph{relaxed microarchitectural isolation} (RMI), a new security property allowing programs to share memory while restricting information leakage to that of non-speculative execution.
To enforce this type of secure speculation, we propose that processors use microarchitectural isolation to restrict the attacker's observations, and a combination of simple mechanisms for controlled speculation and program analysis.
We demonstrate our approach by building an end-to-end FPGA-based prototype running enclave programs with low performance overheads on a multicore out-of-order processor.
Our final platform, Citadel, is a novel design point at the intersection of TEEs and TES defense mechanisms, and a compelling trade-off between performance, usability, and security.

\bibliographystyle{plain}
\bibliography{references}

\appendix

\clearpage
\section{Hardware Software Contracts}
\label{app:hwsfcontract}

We succinctly repeat the definitions and framework for Guarnieri et. al..
Full details can be found in~\cite{guarnieri2021hardware}.
\\

\noindent\textbf{Hardware-Software Contracts}
Contracts define public or exposed states with respect to \emph{programs}.
For a given program $p$ and an initial architectural state $\sigma_0$, $\contract[]{p}{}(\sigma_0)$ is the traces of architectural state when the program is run given the corresponding semantic.
The traces of a contract $\contract[]{p}{})$ capture which architectural state is public and at risk of being exposed by the software when run.
\\

\subsection{Hardware Semantics}
The \emph{adversary model} is modeled as projections of parts of the microarchitectural state.
Given a program $p$, and an architectural state $\sigma$, $\uarch[]{p}{}(\sigma)$ denotes the trace of hardware observations produced in the run of the program on the given microarchitecture.
The hardware semantic of a program $p$ is denoted as $\sigma$, $\uarch[]{p}{}$.
\\

\noindent\textbf{Definition 1} ($\uarch[]{\cdot}{} \vdash \contract[]{\cdot}{}$)
A hardware semantics $\uarch[]{\cdot}{}$ \emph{satisfies a contract} $\contract[]{\cdot}{}$ if, for all programs $p$ and all initial architectural states $\sigma$, $\sigma'$, if $\contract[]{p}{}(\sigma) = \contract[]{p}{}(\sigma')$ then $\uarch[]{p}{}(\sigma) = \uarch[]{p}{}(\sigma')$.

\subsection{Building Blocks for Contracts}
Contracts for secure speculation are defined two fold.
They are composed of a \emph{leakage model} and an \emph{execution model}.
The leakage model describes what is observable by the attacker (e.g. execution trace and trace of memory accesses).
The execution model, on the other hand, models the speculative behavior of the underlying semantic.
\\

\noindent\textbf{Existing Leakage Models}
Several leakage models have been defined and used in previous work.
The \emph{constant-time} model (\sw{ct} for short) leaks the trace of the \texttt{pc} (i.e. the control flow) and the trace all memory accesses (but not their values).
The \emph{architectural} or \emph{sanboxing} model (\sw{arch}) extends the \sw{ct} model to include the values of elements loaded from memory.
\sw{arch} effectively exposes the value of all registers during execution.
The \emph{memory} model (\sw{mem}) only exposes the trace of memory accesses but not the control flow of the program and is sometimes used to model an attacker exploiting cache side-channels.
The more states are exposed to an attacker by a leakage model, the \emph{stronger} it is considered to be (i.e. challenging to defend by software-only measures).
Hence \sw{arch} being the strongest leakage model followed by \sw{ct} and finally \sw{mem}.
\\

\noindent\textbf{Existing Execution Models}
The weaker execution model is \sw{seq}, which represents non-speculative, or sequential execution (one instruction at a time and in-order).
We also have the strongest execution model \sw{spec} which models speculative execution with the following predictors: the pattern history table (PHT), the branch target buffer (BTB) and the return stack buffer (RSB).
\\

\subsection{SW side: Non-Interference Properties}

Work by Cauligi et. al.~\cite{cauligi2022sok} defines direct and relative non-interference properties.
We repeat these definitions here as a reminder.
\\

\noindent\textbf{Direct Non-Interference.}
Program $p$ satisfies \emph{direct non interference} with respect to a given contract $\contract[]{\cdot}{}$ and policy $\pi$ if, for all pair of $\pi$-equivalent initial states $\sigma$ and $\sigma'$, executing $p$ with each initial trace produces the same trace. That is, $p \vdash NI(\pi, \contract[]{\cdot}{})$ is defined as
$$ \forall \sigma, \sigma': \sigma \simeq_{\pi} \sigma' \Rightarrow \contract[]{p}{}(\sigma) = \contract[]{p}{}(\sigma')$$
We elide $\pi$ for brevity and write $p \vdash NI(\contract[]{\cdot}{})$.
\\

\noindent\textbf{Removing Security Policy $\pi$.}
In Cauligi et. al.~\cite{cauligi2022sok}'s definitions of relative non-interference properties all contain a security policy $\pi$, which is used to qualify which part of the initial states $\sigma$ and $\sigma'$ are public or private.
This is redundant for the definition or \emph{relative} non-interference as what is public is already defined by what is leaked in the first contract.
As a result, we elide $\pi$ for brevity.
\\

\noindent\textbf{Relative Non-Interference.}
Program $p$ satisfies \emph{relative non interference} from contract $\contract[a]{\cdot}{seq}$ to $\contract[a]{\cdot}{\beta}$ if: for all pair of initial state $\sigma$ and $\sigma'$, if executing $p$ under $\contract[a]{\cdot}{seq}$ produces equal traces, then executing $p$ under $\contract[a]{\cdot}{\beta}$ produces equal traces.
That is, $p \vdash NI(\contract[a]{\cdot}{seq} \Rightarrow \contract[a]{\cdot}{\beta})$ is defined as
\begin{align*}
\forall \sigma, \sigma': \quad
&\contract[a]{\cdot}{seq}(\sigma) = \contract[a]{\cdot}{seq}(\sigma') \\[1ex]
&\Rightarrow \contract[a]{\cdot}{\beta}(\sigma) = \contract[a]{\cdot}{\beta}(\sigma')
\end{align*}

\section{Burst Mode Security Analysis}
\label{app:proofburstmode}
Given that $\uarch[burst]{\cdot}{} \vdash \contract[shm]{\cdot}{stl}$.
Given that our static program analysis tool \hw{sta} returns \texttt{true} on a program $p$ if $p \vdash NI(\contract[shm]{\cdot}{seq} \Rightarrow \contract[shm]{\cdot}{stl})$, and \texttt{false} otherwise.
Let us prove $\uarch[burst]{p}{sta} \vdash \sw{(}\contract[shm]{\cdot}{seq}$.
\\

Let us take $p$ a program, $\sigma$ and $\sigma'$ some initial state so that
$(\contract[shm]{p}{seq}(\sigma) = \contract[shm]{p}{seq}(\sigma'))$
We have two cases:
\\

\noindent
If $sta(p) = \texttt{true}$, $p \vdash NI(\contract[shm]{\cdot}{seq} \Rightarrow \contract[shm]{\cdot}{stl})$ and $\uarch[burst]{p}{} = \uarch[burst]{p}{sta}$ so:
\begin{align*}
    (\contract[shm]{p}{seq}(\sigma) &= \contract[shm]{p}{seq}(\sigma')) \\[1ex]
    \Rightarrow &(\contract[shm]{p}{stl}(\sigma) = \contract[shm]{p}{stl}(\sigma')) \\[1ex]
    \Rightarrow &  \uarch[burst]{p}{}(\sigma) = \uarch[burst]{p}{}(\sigma') \\[1ex]
    \Rightarrow &  \uarch[burst]{p}{sta}(\sigma) = \uarch[burst]{p}{sta}(\sigma')
\end{align*}

\bigskip
\noindent
If $sta(p) = \texttt{false}$, $\uarch[burst]{p}{sta} = \varnothing$ and:
$$\uarch[burst]{p}{sta}(\sigma) = \uarch[burst]{p}{sta}(\sigma') \text{ is always true.}$$

\bigskip
\noindent
As a result, under our threat model, $\uarch[burst]{p}{sta} \vdash \contract[shm]{\cdot}{seq}$.
\section{Primer on Out-of-Order Processors}
\label{apendixooo}

\begin{figure}[t]
    \centering
    \includegraphics[width=\columnwidth]{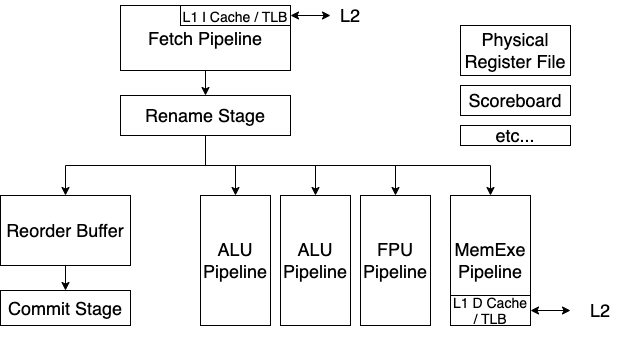}
    \caption{Overview of the Riscy-OOO pipeline.}
    \label{fig:riscypipeline}
\end{figure}

The out-of-order design we consider is split between a front-end (responsible for speculatively fetching new instructions) and a back-end (responsible for executing out-of-order those instructions, emitting the corresponding memory operations, and correcting wrong predictions to redirect the front-end).
An overview of the Riscy-OOO pipeline is provided in Figure \ref{fig:riscypipeline}.
The front-end is irrelevant to this paper. 
The backend starts when an instruction is enqueued both into the reorder buffer (see below) and the correct ALU/FPU/Memory execution pipeline, via each pipeline's reservation station (see below).
Morally, the reorder buffer tracks when an instruction is ready to commit, while the reservation stations track when an instruction is ready to execute.
Together, the reorder buffer and the reservation stations ensure that instructions are executed when their operands become available, as long as it does not contradict the program order. They are also responsible for killing mispredicted instructions and committing right-path instructions.
\\

\noindent\textbf{Reorder Buffer}
Every cycle, the different execution pipelines notify the ROB when an instruction has finished executing, and the instructions' statuses are updated in the ROB when done with execution.
To guarantee correctness, the effect of an instruction is permanently committed only once it reaches the head of the ROB (guaranteeing that all previous instructions in the program have committed).
When an instruction is required to \textit{execute non-speculatively}, the scheduler waits until that instruction reaches the head of the ROB before notifying the reservation station (see below) that this instruction, which is now the oldest in-flight instruction, is ready for execution. 
\\

\noindent\textbf{Reservation Station}
There are multiple reservation stations for different kinds of instruction: ALU, FPU, and the memory instructions.
At each cycle, the entries in the reservation stations are updated to reflect if their operands are ready (that is, if the dependent register values have been computed by previous instructions).
Following this, the scheduler selects, for each reservation station, the oldest instruction that is ready to execute (has all of its required operands and is not), and issues it for execution to its corresponding execution pipeline.

\noindent\textbf{Overview}

\section{Hardware Implementation Details}
\subsection{Safe Mode}
\label{app:safemode}
In the baseline out-of-order processor we are hardening, memory instructions are decoded, renamed, and then simultaneously queued in the reorder buffer (ROB) and the memory reservation station (\MemRS{}) (see Appendix~\ref{apendixooo}).
When an instruction's operands are ready, it is dispatched to the memory execution pipeline (\MemExePipeline{}).
Safe mode identifies instructions that access untrusted memory early in the \MemExePipeline{} and delay the translation and processing of the memory accesses until the instruction is no longer speculative, i.e.,  when it reaches the head of the ROB.
These changes require modification of \MemRS{} and \MemExePipeline{}. 
\\

\noindent\textbf{Modifying the Memory Execution Pipeline}
Upon entry into \MemExePipeline{}, checks are performed on memory instructions to decide if the memory operation is safe to translate and then issue or not.
It is important to perform these checks very early as
the next steps in the \MemExePipeline{} touch shared microarchitecture (through address translation and cache hierarchy, for instance) potentially transmitting information to an attacker.
A memory access is deemed safe if it accesses trusted memory OR if it is non-speculative.
Because the enclave trusted memory range is defined in virtual memory, this check can be performed as soon as operands (i.e. virtual address) are available.
If an instruction tries to access untrusted or shared memory, we check that it is non-speculative by conservatively checking if it has reached the head of the ROB.
If the instruction is safe, it is dispatched to the next stage of the \MemExePipeline{}.
If it is not safe, the instruction is squashed and a signal is sent to the \MemRS{} indicating that it should re-dispatch it once it reaches the head of the ROB.
\\

\noindent\textbf{Decorrelating Request Dispatching and Request Dequeuing}
Safe mode requires the \MemRS{} to retain the accesses to shared memory in order to re-dispatch them once they become non-speculative.
The original design performs a \MemRS{} data dispatch and dequeue simultaneously once the request enters the \MemExePipeline{}. 
These two operations now need to occur at different stages in the pipeline: data is first dispatched from the \MemRS{} and is dequeued in the translation execution stage if and only if the request is determined to be secure. 
We modify the interface between \MemRS{} and \MemExePipeline{} to provide the correct interfaces and logic.
This includes augmenting the request data with the corresponding \MemRS{} entry index.
\\

\noindent\textbf{Modifying the Memory Reservation Station}
Our mechanism requires \MemRS{} being able to re-dispatch some instructions once they become non-speculative: untrusted memory loads will indeed typically need to get dispatched twice.
First, several new status bits are added for each \MemRS{} entry, specifically to track if a) a given request is still to be dispatched for the first time, and b) if a given request needs to be re-dispatched. 
The first bit (\textit{to\_dispatch}) is necessary since the entries can now still reside in the \MemRS{} in the time between being dispatched and dequeued (or signaled for redispatch) and should not be selected for dispatch again.
The second bit (\textit{to\_redispatch}) is set by \MemExePipeline{} when it squashes a speculative insecure request. 
We additionally add logic to compare each \MemRS{} entry to the entry at the head of the ROB, and raise the entry's \textit{dispatch\_ready} signal on a match.
Finally, the dispatch logic is modified to use the new status bits. Rather than simply checking for the oldest valid entry whose arguments are ready, an additional condition is added to check if either \textit{to\_dispatch} is set for the entry or \textit{to\_redispatch} and \textit{redispatch\_ready} are set for the given entry.
\\

\noindent\textbf{Disabling the Mechanism for the OS}
Blocking speculation on memory access can have a significant performance impact.
We want to make sure that the OS is still allowed to speculate on its own memory if it chooses not to use our trusted memory mechanism.
As a result, we disable our mechanism when the enclave virtual address range is set to a specific empty range.
Note that the OS could always be modified to take advantage of our trusted memory mechanism and protect itself from speculative side-channel attacks.

\subsection{Reconfigurable LLC Partitioning}
\label{app:dynamiccache}
Static partitioning (as proposed in MI6) is impractical as each of the 64 DRAM regions would be allocated a 16KB cache slice in our 1MB LLC.
This is even smaller than the 64KB necessary for the inclusion of the cores' private L1 caches (D1 and I1), which is dramatic for performance.
Using way partitioning \cite{kiriansky2018dawg} is not possible as we need to instantiate 64 regions in the cache, but the LLC has only 16 ways, and Ironhide's \cite{ironhide} strategy assumes limited reconfiguration.
\\

\noindent\textbf{Hardware-Software Co-Design}
The key is to rely on the (trusted) SM to do most of the heavy lifting in terms of bookkeeping and checking for correct configuration and security.
At the hardware level, we can now design a new, simple, and practical reconfigurable LLC partitioning scheme that only requires minimal hardware changes.
\\

\noindent\textbf{Configuration Protocol}
We build a mechanism for the OS to reconfigure the LLC partitions to improve enclave and other program's performance.
This costly operation is only required when changing the LLC layout, which happens only before launching an application or an enclave with an important memory footprint.
The OS requests a reconfiguration through an SM call while no enclaves are running.
The SM will first check that the new partitioning is valid, doesn't affect regions assigned to the SM.
It then waits for other cores to flush their L1 caches and to enter a blocked state in the SM, making sure no requests will be sent to the modified regions during reconfiguration (which could put the cache in an incoherent state).
The SM will then interact with the hardware to flush the LLC slices that are modified and set the LLC in its new configuration.
\\

\noindent\textbf{Microarchitectural Implementation}
\begin{figure}[t]
    \centering
    \includegraphics[width=\columnwidth]{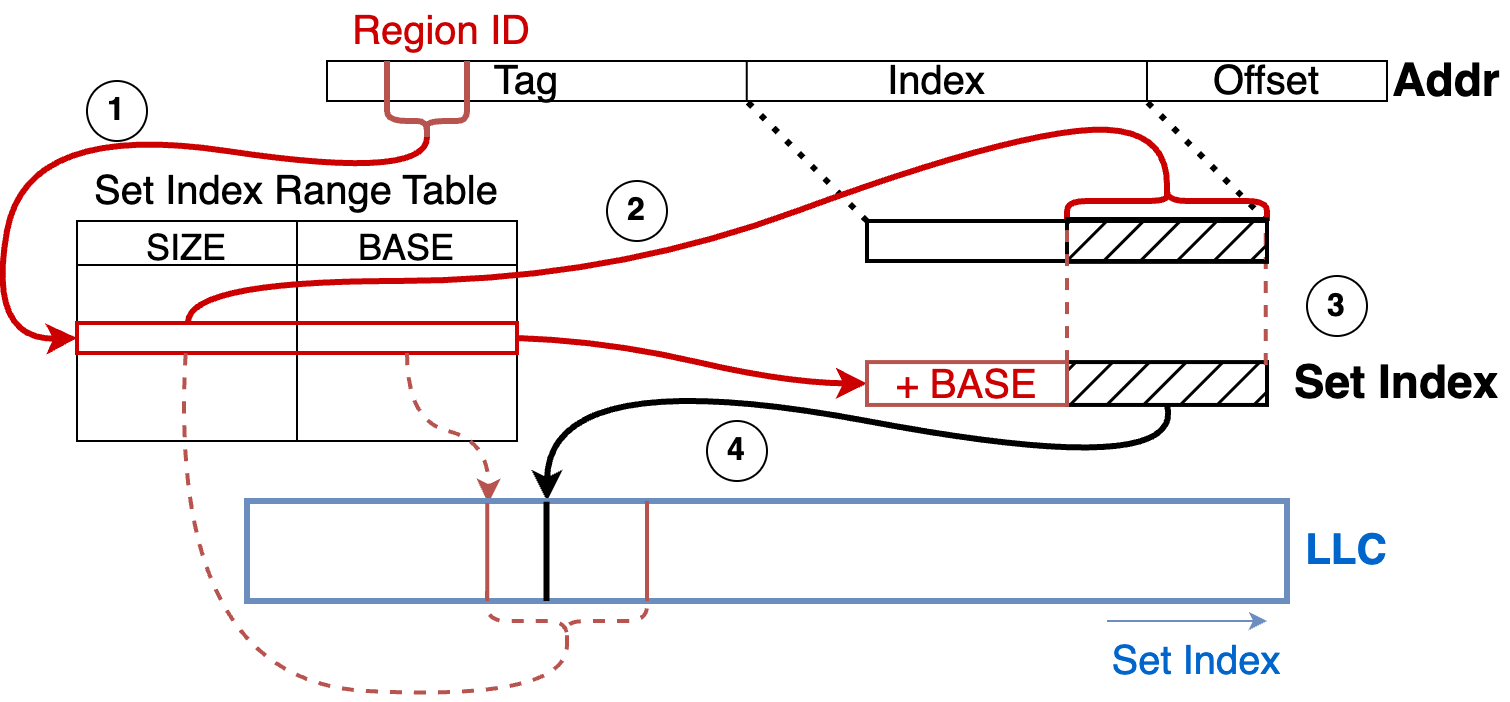}
    \caption{LLC set indexing scheme. The function takes an address as an input and output a set index in the LLC slice allocated to its memory region.}
    \label{fig:dynamicpartitioning}
\end{figure}
The configuration for each LLC slice (base and bound of the set-index range) is stored in a big register (1216 bits) called the set-index range table.
The SM can modify the entries of the set-index range table by sending MMIO requests to the cache, which is seen as a new device.
When the LLC receives a request, it will compute the new set index for the corresponding address.
We represent the transformation used in Figure \ref{fig:dynamicpartitioning}.
First, we extract a region ID from the address upper bits and obtain the corresponding index range base and size by looking up the set-index range table \circled{1}.
We then extract the original set index of the address, located in the middle of the address, and tweak it.
We compute the original set-index value modulo the size of the index range \circled{2} and add the base index \circled{3}.
This effectively restricts the new set index to the corresponding index range, partitioning the LLC \circled{4}.
Note that we must extend the length of the tags used by the LLC to include the entire original index in case an index range is of size 1 and all addresses from the same region map to the same set.
\\

\subsection{LLC fine-grain flush}
\label{app:llcflush}
To go hand-in-hand with our reconfigurable LLC partitioning, we also want to be able to selectively flush parts of the LLC.
We create a \emph{Zero Device} in the upper physical address space. 
It is implemented as a simple piece of logic connected to the cache hierarchy at the same level as the DRAM, but that always sends back zeros.
The trick is that the address space covered by the zero device has the same size as the DRAM and is located at addresses that will collide with DRAM addresses in the cache.
Once the SM has placed the machine in a correct waiting state (where all the cores have flushed their L1 and are waiting in the SM), it only needs to access an eviction set in the zero-device address space to flush the cache slice.
This means that flushing a small LLC slice is much faster than flushing a big one.
We look at the performance of our flushing mechanism in the evaluation section (see \ref{sec:evalllcpartitioning}).

\subsection{Relaxed Flushing Policies}
\label{app:relaxedflushing}
The enclave and some parts of the mini-SM do not have secrets to keep from each other.
Specifically, the mini-SM never manipulates sensitive data.
As a result, the mini-SM does not need to flush microarchitectural state when context switching to enclaves.
This reduces the cost of mini-SM calls and opens the door to the use of frequent software or timer interrupts.
To disable flushing selectively, we modify the hardware to deactivate systematic flushing and exposed the SM a register to trigger the flush.
We write the SM to trigger the flushing mechanism on required context switches, but not on most traps and SM-calls from the enclave.
\\

\noindent\textbf{Evaluation}
\label{app:evalrealxedflush}
\begin{figure}[t]
\centering
\includegraphics[width=\columnwidth]{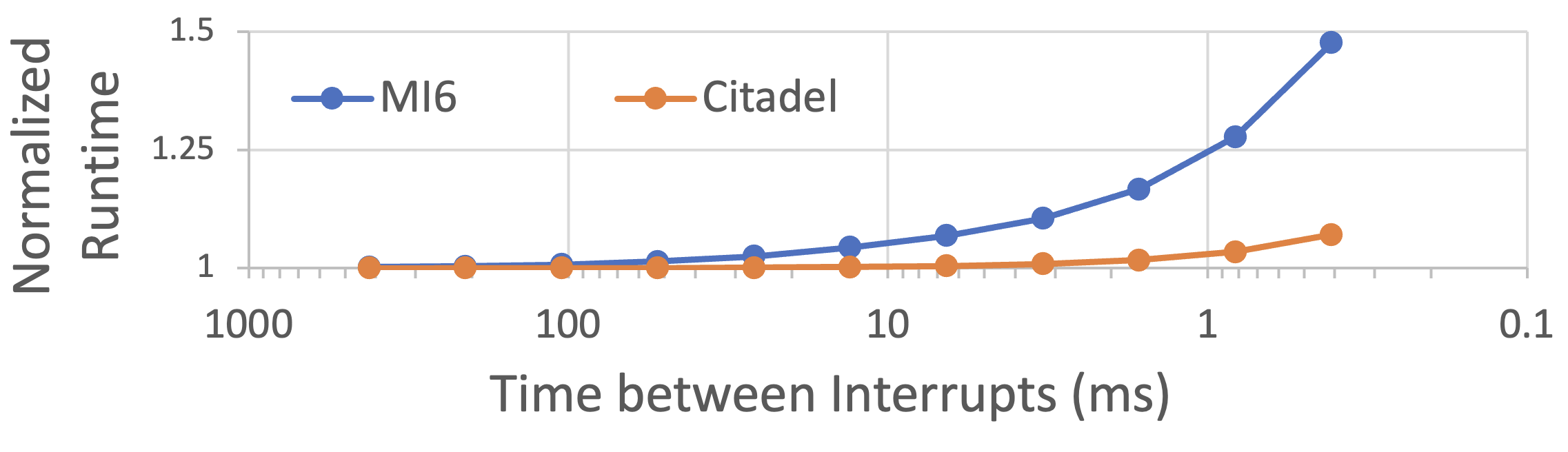}
\caption{Normalized Execution Time for Coremark in the presence of periodic timer interrupts. MI6 flushed on every interrupt; Citadel does not.}
\label{fig:coremark_timer}
\end{figure}
To evaluate the performance of the relaxed flushing policy, we run Coremark while triggering periodic timer interrupts at different time intervals to model frequent traps.
We do not perform any computation inside the trap handler and only measure the overhead due to context switching.
The results are provided in Figure \ref{fig:coremark_timer}.
By testing a range of interrupt periods, we confirm that flushing on each trap like in MI6 results in significant overheads.
However, the cost is greatly reduced when using the relaxed flushing policy, with a maximum of 7\% overhead measured for interrupts every 0.4 ms.

\subsection{Miscellaneous}
We statically partition MSHRs between two cores and place a fair round-robin arbiter at the entry of the LLC.
Because these two mechanisms only marginally affect the latency of the LLC for a 2-core design, we observe negligible ($\approx$ 0.1\%) performance impact.

We also expose controls for speculation and predictors to the software through Control Status Registers (CSRs).
Specifically, a program is able to deactivate speculation all together (every instruction waits to reach the head of the ROB), delay the issuing of shared memory accesses to the head of the ROB, deactivate the training of predictors (branch predictors and return address stack) and/or their use (and then it falls back to the predictor $pc + 4$, or straightline speculation).

\section{Software Infrastructure}
\label{app:softwareinfrastructure}

\subsection{Security Monitor}
The SM does not allocate any resources, this task is left to the OS, but it will interpose between each security-sensitive operation (like a resource allocation) and check that the operating system performed the task without violating the high-level security policies (e.g., it is allocating a memory region to an existing enclave).
Its API is called by the OS or an enclave to request ownership of new memory regions, create enclaves, or perform any other operations where it is needed.
As a result, the SM is mostly responsible for bookkeeping operations, performing security checks, and setting up the hardware mechanisms to enforce isolation between the running security domains.
It will not only systematically intervene on any security-sensitive operations, but will also play the role of a small kernel and build the bridge between the OS and the hardware on some non-security-sensitive operations like sending inter-process interrupts (IPI).
The OS and the enclaves interact with the SM through ecalls.
A more detailed description of the SM data structures and API, can be found in \cite{lebedev2019sanctorum}.

\subsection{Boot Sequence}
\noindent\textbf{Secure Bootloader}
To boot an OS such as Linux on top of our SM, several steps are necessary.
First, our Secure Bootloader will be in charge of the first booting stage.
While other cores are waiting,
core 0 will load, measure (hash) the SM binary, and use the SM measurement to generate the SM key pair. 
Once this is done, the Bootloader will use its private keys to sign the SM measurement and endorse the SM public key, creating the first certificate used for secure boot or remote attestation.
Finally, it will copy the SM key pair to SM memory along with the endorsing certificate.
For more details on secure boot, see~\cite{lebedev2018secure}.
It will then wake up the other cores using interprocess interrupts and have them all jump to the SM entry point.
\\

\noindent\textbf{SM Boot}
Now that the SM has been attested and is running, it can start initializing the hardware, set up the data structures required to later perform its duties, and set up the machine is in such a state so that Linux can properly boot in the next stage.
The SM proceeds as follows: each core will clean up its register file and set up a stack.
They will all initialize some of their special registers.
In particular, they will immediately set up a trap handler to be able to catch exceptions and enable software interrupts.
From there, all cores go to sleep except core 0.
Core 0 will set up all private SM data structures that keep track of the ownership of resources.
It will also walk the device tree and set up kernel-related mechanisms such as the clock and interprocess interrupts.
The other cores will then be awakened.
The SM will set up memory protection through the specific registers.
In particular, some memory checks are performed during address translation.
However, Linux uses physical addresses to boot.
To solve this issue, the SM sets up an identity page table, so the virtual address space exposed to Linux is a direct mapping to the physical address space.
Each OS memory access will then go through address translation and memory security checks.
Finally, the SM will have all the cores jumping to the payload entry point at a similar time; in our case, the payload is the Linux image.
\\

\noindent\textbf{Linux Booting and SM Kernel Module} 
We boot Linux using the Spinwait method \cite{dabbelt2017all}.
Once Linux boots, an SM Kernel Module will be loaded.
The SM kernel module is registered as a device in Linux that is accessible to applications through a pseudo-file \texttt{/dev/security\_monitor}.

\subsection{Running an Enclave}

\subsubsection{Key Software Elements}
\noindent\textbf{The SM Kernel Module}
The main SM should not possibly be made unavailable to the OS by a user-level program.
Furthermore, no resource should be allocated to a user-level program without the OS approval.
As a result, to create an enclave or perform other operations,
a user application needs to interact with the SM through the SM Kernel Module.
Because the SM Kernel Module is part of the OS, it can then refuse requests if needed.
\\

\noindent\textbf{Mini-SM}
Because the SM is a shared resource, its might leak sensitive information. For instance, an enclave might call the SM at particular times and in a pattern that might leak sensitive information regarding the enclave's secrets.
As a result, each enclave uses its own private copy of the SM.
This "mini-SM" is smaller than the one accessible to the OS and only contains the code corresponding to the subset of the SM API accessible by the enclave.
It also uses its own private stack.
This mini-SM is located in private enclave memory (not accessible nor visible by the OS) and protected using physical memory isolation (hence it is neither modifiable nor accessible by the enclave).
The enclave interacts with its mini-SM through the ecall API.
Note that some of these ecalls might touch on shared global SM state and locks.
As a result, an enclave should be careful when interacting with the SM and be aware that it might still leak timing information to other programs.
We emphasize that an enclave binary does not go through the SM Kernel Module and directly interacts with its mini-SM through ecalls.
\\

\noindent\textbf{User-Level Application}
The user-level application runs on top of the OS.
It does not benefit from any integrity or privacy guarantees, unlike its enclave counterpart.
On the other hand, it has access to a rich interface with the outside world (like the file system or the network) through all the usual OS services implemented through system calls.
Even if not required for an enclave to work, this makes the user-level application a good ally for an enclave program.
It can, for instance, fetch encrypted private data and forward it to the enclave or perform many other non-security-sensitive tasks.

\subsubsection{Launching An Enclave}
\label{app:launchingprocess}
\noindent\textbf{Allocating Memory}
The user-level application first loads the enclave binary from the file system to user memory.
It will also ask the OS to map the virtual addresses used by the enclave for shared memory to physical memory.
The application should also make sure not to allocate any shared objects within the enclave's virtual memory range.
The application then opens the kernel module device and sends a request through an \texttt{iotcl} system call to request the SM Kernel Module to launch the enclave binary.
\\

\noindent\textbf{Setting Up the Enclave}
Once called, the SM Kernel Module launches a series of ecalls to the SM to setup the enclave.
First, it moves the enclave binary from user memory to kernel memory.
It then allocates two physical memory regions and performs several SM calls so the SM can use one of the regions to store metadata and the second region can be given to the enclave.
After the SM has set up the metadata region, the enclave's metadata can be created and populated.
During the enclave setup process, the SM makes sure to add several arguments used to set up the enclave to the enclave's measurement (see Section \ref{app:attestation} for more details).
The kernel module then asks the SM to copy a version of the mini-SM in enclave memory, initializes the enclave's page table, and loads the enclave's binary while populating the enclave's page table with entries.
The kernel module asks the SM to create one or several threads and gives them to the enclave. 
The threads are set up to start executing at the right location and with the right stack pointer.
\\

\noindent\textbf{Timer Limit}
We also added a new value \texttt{timer\_limit} to each thread.
This value is used to avoid timing attacks on enclaves, especially ones exploiting malicious inputs to leak enclave secrets through the enclave completion time.
\texttt{timer\_limit} will represent the maximum amount of time the enclave's thread is allowed to execute.
During enclave launch, the hardware timers are set so the enclave will be exited after the corresponding amount of time has elapsed.
This way, if an unrecoverable exception occurs, the enclave will not immediately exit, but will simply wait for its timer to expire.
This prevents the enclave from leaking the timing at which exceptions occurred.
\\

\noindent\textbf{Sealing the Enclave}
The enclave can now be sealed, its measurement is finalized and used to generate a key pair.
The SM commits to the enclave's measurement and public key using its own signing key.
The enclave is now ready and the module can send a final ecall to the SM to enter the enclave.
\\

\noindent\textbf{Enclave Execution}
The enclave now executes on its dedicated cores and can communicate with its parent-level user-application using shared memory.
When the enclave wants to exit, it does so by sending a ecall directly to the mini-SM.
Upon return, the kernel module asks the SM to delete the threads and the enclave's metadata; it then reclaims the memory regions and gives them back to the OS.
Finally, the kernel module returns and forwards the enclave's return value to the user-level application.

\subsection{Enclave's Attestation Mechanism}
\label{app:attestation}
Remote attestation makes it possible for a remote user to receive and check a proof that convinces them of the integrity of an enclave's initial state, given trust in the Secure Bootloader and the SM's code.
Our remote attestation protocol is based on \cite{lebedev2018secure}.
The attestation proof is composed of two certificates, one ensuring the integrity of the SM and one ensuring the integrity of the secure enclave.
\\

\noindent\textbf{SM Attestation}
First, during secure boot, our (trusted) Secure Bootloader will measure the SM binary using a cryptographic hash function (SHA3) along with the device's secret, to derive the SM signing keys.
It will then generate a certificate using a public-key signature algorithm (Ed25519) with a key pair generated from a root device secret (generated on-chip, for example, using a PUF \cite{herder2014physical}) by signing the SM's measurement and its public key.
The SM certificate, the SM keys and the Bootloader public key are copied to SM memory by the Bootloader.
The SM certificate and the Bootloader public key are made available to the OS and to enclaves through the SM API.
The bootloader is then responsible for setting up the SM stack correctly and entering the SM at the right address.
\\

\noindent\textbf{Enclave Attestation}
During the enclave's setup, the SM will create the enclave's unique measurement by hashing (with SHA3) several arguments used to set up the enclave initial state.
In particular, it will include the enclave's trusted virtual address range, the enclave's binary along with its virtual memory mapping, the mini-SM binary and for each thread, the entry \texttt{PC} and \texttt{SP}, the value of \texttt{timer\_limit}, and a representation of which exceptions and interrupts are enabled and which ones are directly delegated to the enclave.
Because our SM is trusted, and the Ed25519 algorithms are implemented in constant time and protected against speculative and non-speculative side-channel attacks (speculation is disabled inside of our SM), we modify the original Sanctum scheme \cite{costan2016sanctum} and do {\em not} rely on an attestation enclave to sign the enclave's measurement and generate the enclave's certificate.

\begin{figure}[t]
\centering
\includegraphics[width=\columnwidth]{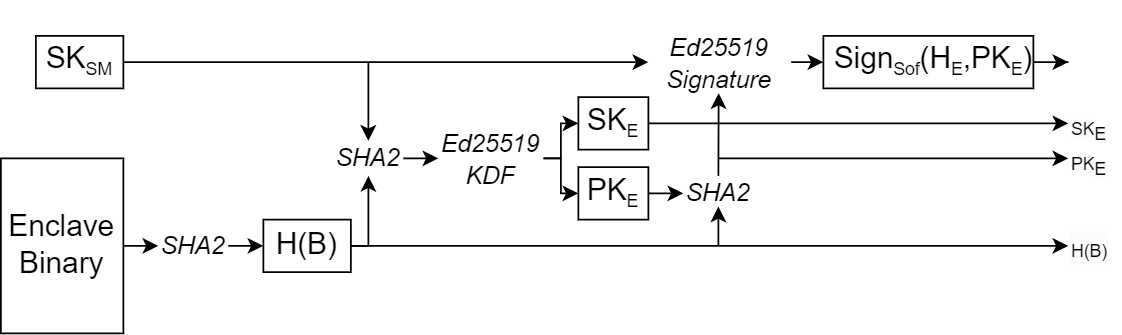}
\caption{Key Derivation and Endorsement}
\label{fig:keyderiv}
\end{figure}

Instead, after the enclave's measurement has been completed, the SM will use it, along with a hash of its secret key, to derive a Ed25519 key pair for the enclave following a similar process to the Bootloader (see Figure \ref{fig:keyderiv}).
It will then generate an attestation certificate by signing the enclave's measurement and the enclave public key.
The enclave measurement, certificate and keys are copied to the enclave's metadata region, which is private to the SM.
\\

\noindent\textbf{Extending the SM Attestation API}
We introduce two new SM API calls.
One so an enclave can retrieve its own keys and attestation certificate (no one else can retrieve its private key).
And a second so anyone can retrieve any enclave attestation certificate (which contains its public key).
The enclave key pair can then be used to sign messages or to perform key exchanges and create secure communication channels \cite{bernstein2012high}.

This solution has several advantages compared to the Sanctum attestation enclave scheme.
The enclave does not need to rely on another enclave to perform its attestation.
This avoids several communication, concurrency and availability issues, for instance, how an enclave finds or wakes up the attestation enclave.
This also means we don't need to share the attestation enclave with other potentially untrusted enclaves.
Additionally, the key-derivation scheme is now deterministic.
This means that the same enclave on the same machine will be endorsed with the same keys given an identical SM and a fixed device secret.
This makes it possible for a remote user to process shared secrets with the enclave and prepare and encrypt data for the enclave without the enclave actively running.
This also makes it possible for an enclave to derive deterministic keys for long-term secure data storage that can be recovered even after being offline.
\\

\noindent\textbf{Position-Independent Enclave Measurement}
For our attestation mechanism to be practical and our enclave's key derivation to be deterministic, it is important for an enclave's measurement to be independent of the enclave's location in physical memory.
It would be impractical for a remote verifier to keep a copy of a hash for each possible enclave's placement in physical memory (given that an enclave could own any subset of the 64 physical memory regions, which would be at worst $2^{64}$ elements).
We need to tweak Sanctum's mechanism to properly enforce that property.
This property is enforced two ways.
\\
\noindent\textbf{Position-Independent Mini-SM}
First, for the mini-SM, its code must be completely position-independent. 
This is achieved at compilation time by using specific flags. 
Additionally, the mini-SM is only accessed through traps.
The entry into the trap handler and the mini-SM stack must be fixed relative to the mini-SM binary.
Because these values are physical addresses and cannot be added to the enclave's measurement, the SM is trusted to set up the trap handler and the stack correctly.
\\

\noindent\textbf{Mapping Virtual Addresses to Bytes}
Second, for the enclave's binary, we can rely on virtual memory to enforce that the enclave's code will always use the same addresses independently of its position in physical memory.
The only challenge left is how to provide a measurement that reflects the virtual address mapping without including any physical addresses. Instead of measuring the two mappings virtual-address-to-physical-address and physical-address-to-byte-content (for instance, by measuring the page table and measuring the physical memory content and its addresses), we will directly enforce and measure the mapping from virtual-addresses-to-byte-content.
When loading a page from the enclave's binary, the SM will be given, by the OS, the physical address at which the enclave page should be copied, the virtual address at which the page should be mapped and a pointer to the enclave's binary page in OS memory. 
It will then 1) copy the page to the destination physical address, 2) set up the enclave's page table to enforce the right mapping, and 3) only add to the enclave's measurement the hash of the page content, the hash of the virtual address it was mapped to, and the hash of the page table entry flags used.
As a result, it is enough for a remote verifier that trusts the SM to 1) check the SM certificate and 2) check the enclave's measurement to correctly identify the enclave.
\\

\noindent\textbf{Extension to Trusted VMs}
Note that for an enclave that manages its own virtual memory, we would need to play other tricks to make its measurement position independent.
Indeed, the enclave would have somehow to learn about its position in physical memory, by hardcoding it (and hence fixing it) it or by reverse-engineering the page table provided by the SM.
Hence, porting a complete operating system inside an enclave would require modifications to achieve a realistic end-to-end attestation solution.

\subsection{Building an Enclave Application}

\noindent\textbf{Enclave Runtime}
The enclave binary or enclave application is the code that will run inside the enclave.
In should really be seen as a binary as, at first, the execution environment made available to the enclave is close to bare metal.
Indeed, to enforce the integrity and privacy of the enclave, its private memory and its execution cores are architecturally and microarchitecturally isolated from the OS.
This means, for instance, that the enclave cannot directly interact with the OS through the usual syscalls.
As a result, an enclave binary needs to include its own libraries.
Recall also that the OS does not have access to the enclave's private memory.
That means the enclave doesn't have access to the OS resource management and needs to manage resources (e.g., memory, threads, etc.) on its own.

However, the enclave has access to its mini-SM through the \verb|ecall| interface and can perform operations such as freeing a memory region, setting up timer interrupts, or exiting.
The enclave also has access to shared memory: it can read and write the data that is made available to it by the OS.
In our implementation, the enclave can access the address space of the user-level application that launched it.
This makes our enclaves expressive and allows them to run rich applications such as cryptographic libraries, language run-times, or ML inference.
Inputs can be shipped directly with the enclave binary or communicated to the enclave using shared memory.
To protect the integrity and privacy of the input data sent to an enclave, a remote user can establish a shared secret with the enclave using the enclave's public key and encrypt the data before sending them to the enclave (see Section \ref{app:attestation}).

\subsection{Debugging Infrastructure}
\noindent\textbf{Debugging an Application}
For early software development and debugging, we developed a QEMU target that matches our processor.
A programmer can easily run their applications on several cores on top of QEMU without requiring to run the enclave from within Linux.
From there, the application can be easily debugged using GDB.
We also make it possible to compile the mini-SM with specific debugging flags to allow an enclave to (insecurely) access the console and to have the SM printing extra debugging messages, such as information regarding enclave exceptions.
Once Linux has been booted, applications can be debugged on FPGA using standard Linux debugging tools.
\\

\noindent\textbf{Debugging the Enclave Platform}
Our QEMU simulator can also be used to debug other elements of the software stack including the Secure Bootloader, the Security Monitor, Linux itself, but also the SM Kernel Module, enclave binaries, and user-level applications.
In addition, the hardware can be compiled to run in Verilator, a cycle-accurate simulator.
This can be especially handy for debugging new hardware features or bugs that arise only on the FPGA.
We also developed a series of low-level hardware tests to check the memory protection primitives such as the dual-page-table mechanism.
To conclude, the debugging infrastructure makes it possible to efficiently implement new mechanisms whether they are implemented in hardware, software or a combination of both.

\end{document}